\documentclass[12]{article}
\usepackage{graphics,graphicx,amsmath,amsbsy,amssymb,color}

\usepackage{arxiv}

\newcommand{\JJ}{{\boldmath \mbox{$J$}}}

\newcommand{\uu}{{\boldmath \mbox{$u$}}}
\newcommand{\bb}{{\boldmath \mbox{$b$}}}
\newcommand{\ww}{{\boldmath \mbox{$w$}}}

\newcommand{\WW}{{\boldmath \mbox{$W$}}}
\newcommand{\XX}{{\boldmath \mbox{$X$}}}

\newcommand{\rr}{{\boldmath \mbox{$r$}}}

\newcommand{\pp}{{\boldmath \mbox{$p$}}}
\newcommand{\vv}{{\boldmath \mbox{$v$}}}

\newcommand{\nn}{{\boldmath \mbox{$n$}}}

\newlength{\defbaselineskip}
\newcommand{\setlinespacing}[1]%
           {\setlength{\baselineskip}{#1 \defbaselineskip}}

\DeclareMathOperator{\arcsinh}{arcsinh}

\title{On the role of geometry in statistical mechanics and thermodynamics II: Thermodynamic perspective}

\author{O{\u g}ul Esen\\
Department of Mathematics, Gebze Technical University,\\
41400 Gebze, Kocaeli, Turkey\\
\And Miroslav Grmela \\
\'{E}cole Polytechnique de Montr\'{e}al,
  C.P.6079 suc. Centre-ville,\\
 Montr\'{e}al, H3C 3A7,  Qu\'{e}bec, Canada\\
Corresponding author: miroslav.grmela@polymtl.ca\\
\And and Michal Pavelka \\
 Mathematical Institute, Faculty of Mathematics, Charles University,\\
 Sokolovsk\'{a} 83, 18675 Prague, Czech Republic
}

\newcommand{\shorttitle}{Thermodynamic perspective}

\begin{document}

\maketitle

\begin{abstract}

The General Equation for Non-Equilibrium Reversible-Irreversible Coupling (GENERIC) provides structure of mesoscopic multiscale dynamics that guarantees emergence of equilibrium states.
Similarly, a lift of the GENERIC structure to iterated cotangent bundles, called a rate GENERIC, guarantees emergence of the vector fields that generate the approach to equilibrium.
Moreover, the rate GENERIC structure also extends Onsager's variational principle.
The MaxEnt  (Maximum Entropy) principle  in the GENERIC structure becomes  the  Onsager variational  principle in the rate GENERIC structure. In the absence of external forces, the
rate entropy  is a potential that is
closely related to the entropy production.
In the presence of external forces when the entropy does not exist, the rate entropy still exists. While the entropy at the conclusion of the GENERIC time evolution  gives rise to equilibrium thermodynamics, the rate entropy at the conclusion of the rate GENERIC time evolution gives rise to rate thermodynamics.
Both GENERIC and  rate GENERIC structures are put into the  geometrical framework in the first paper of this series. The rate GENERIC is also shown to be  related to Grad's hierarchy analysis   of reductions of the Boltzmann equation.
Chemical kinetics and kinetic theory provide illustrative examples. We introduce  rate GENERIC extensions (and thus also Onsager-variational-principle formulations) of both chemical kinetics and the Boltzmann kinetic theory.

\end{abstract}


\section{Introduction}

Externally unforced and internally unconstrained macroscopic systems are seen in experimental observations to approach equilibrium states at which their behavior is well described by equilibrium thermodynamics. The time evolution that explicitly displays  such approach has first emerged in Boltzmann's \cite{Boltzmann-vorlesungen} investigations of the time evolution that takes place in ideal gases. State variables of the equilibrium thermodynamics as well as the fundamental thermodynamic relation arise as a pattern in the phase portrait (i.e. collection of all solutions)  of the Boltzmann equation.
Boltzmann's insight has served as a guide for the same type of investigations in   a large family of   macroscopic physical systems. For instance hydrodynamics (that uses  hydrodynamic fields instead of the one particle distribution function as state variables)  together with the assumption of local equilibrium provides another  setting  for investigating  the emergence equilibrium thermodynamics in the time evolution of a large family of fluids. The mathematical structure that is essential for the emergence  of equilibrium states in phase portraits of mesoscopic time evolution equations has been collected in an abstract GENERIC equation. The original Boltzmann equation as well as hydrodynamic and other mesoscopic time evolution equations whose solutions show the  emergence of equilibrium thermodynamics are  all its particular realizations.

The state variable in the abstract GENERIC equation is left unspecified. The  vector field that generates the mesoscopic time evolution is the sum of the Hamiltonian part (representing mechanics) and the gradient part (representing thermodynamics). The building blocks in which the individual features of the macroscopic systems under consideration are expressed are  three  potentials (having the physical interpretation of energy, entropy and number of moles) and two geometrical structures (one transforming the gradient of energy to the Hamiltonian  vector field and the other transforming the gradient of entropy into the gradient vector field). The GENERIC  time evolution equation represents also a dynamical formulation of the Maximum Entropy principle (MaxEnt). The entropy is maximized by following the time evolution generated by the GENERIC vector field.

The abstract GENERIC  structure has been found  useful in modeling the time evolution of complex physical systems \cite{adv} and in data driven modeling \cite{paco-miroslav}. In both cases, the GENERIC  structure brings an organization. After selecting the level of description by specifying the state variables,
the GENERIC  building blocs can be searched  separately. If they satisfy all the required properties  and if they are put correctly in the places reserved for them, then model predictions are guaranteed to approach a state of thermodynamic equilibrium, described by classical equilibrium thermodynamics. In the case of complex fluids, the building blocks are found by using a physical insight into the systems under investigation, while in the case of data-driven modeling, the blocks are found from collected data.

In the first paper \cite{OMM-I} of this series of two papers, we ask the question of what is an appropriate geometrical environment for GENERIC. 
The problem is that Hamiltonian mechanics is usually represented in different geometric setting than gradient dynamics, and thus it is not clear in which setting their sum (GENERIC) should be formulated.
Robert Hermann \cite{hermann} and Ryszard Mrugala \cite{Mrugala-geometrical} have realized that the Gibbs formulation of equilibrium thermodynamics \cite{gibbscw} finds its most natural geometrical environment in the contact geometry. Indeed, the Legendre  transformations,  that are of fundamental importance in the equilibrium thermodynamics,  preserve the contact structure.
We have extended Hermann's observation to dynamics. In the "geometrization" of dynamics, the features and concepts that are arising  in the investigation of the phase portrait are taking the guise of geometrical structures. In this way they become manifestly displayed.
In one of the formulations inspired by geometry,
the GENERIC time evolution is shown to be a contact structure preserving time evolution in $T^*M\times \mathbb{R}$ (where $M$ is the state space and $T^*M$ is its cotangent bundle) restricted to one of its Legendre submanifolds.
In this second paper, we investigate  the physical interpretation of the geometrical formulations. We also show (in Section \ref{sectiongeneric}) how the  equilibrium thermodynamics emerges as a pattern in the phase portrait generated by GENERIC.

In Section \ref{sectionRgeneric}, we investigate reductions of detailed GENERIC descriptions to less detailed ones. Rayleigh \cite{Rayleigh} followed by  Onsager \cite{onsager1930}, Prigogone \cite{prigogine-tip}, Gyarmati \cite{gyarmati}, and more recently by Doi \cite{doi-onsager}, Rajagopal and Srinivasa \cite{raja-sri}, and Guo and Hou \cite{thermomass}  have  replaced entropy with a dissipation potential in the role of the potential that is extremized in the reduction process. Instead of maximizing the entropy subjected to constraints (that are constants of motion in the approach to equilibrium), it is the dissipation potential that is extremized subjected to constraints (that represent fluxes in the reduced theory). This variation principle
is called Rayleigh's
principle of the least energy dissipation  or in its extended form  Onsager's variational principle,
\begin{equation}
\frac{\delta D}{\delta \dot{q}} = \frac{\delta F}{\delta q},
\end{equation}
where $q$ is a set of state variables, $D$ a dissipation function and $F$ a free energy \cite{onsager1931,landau5}.
The Onsager's principle is usually considered as unrelated to GENERIC, without geometrical interpretation, without clear reasoning of the minimization, and without direct relation to maximization of entropy.
We, on the other hand, get a dynamic version of Onsager principle by lifting GENERIC, keeping the static version as a consequence of the dynamic one. We show that  the physics behind it is the same as the physics behind the MaxEnt, except that MaxEnt is the reduction to equilibrium thermodynamics while the Onsager principle reduces to a less detailed dynamical theory.
In this perspective, the reduced theory is not the equilibrium thermodynamics, but another dynamical theory involving less details.


In this paper we provide a relation between GENERIC and Onsager's variational principle, giving the latter a geometric meaning. We derive Onsager's variational principle from GENERIC in two ways. First, in a bottom up derivation we lift GENERIC to iterated cotangent bundles. Second is the top down derivation in which we begin with GENERIC representing a more detailed theory, reformulate it into a hierarchical form, and close the hierarchy. The Onsager variational principle arises in  the closure. Both derivations enrich  Onsager's variational principle by providing  the time evolution making the extremization and providing, in addition to the reduced dynamics, the  rate fundamental thermodynamic relation inherited from the process of extremization. The latter  relation is analogous to the fundamental thermodynamic relation inherited from the maximization of the entropy in the MaxEnt principle. The bottom up derivation has started already in Part I \cite{OMM-I}, where we called the lifted GENERIC a rate GENERIC. We keep this terminology and use systematically the adjective "rate" to make a distinction between the investigation of the approach to equilibrium initiated by Boltzmann (and put into an abstract form in GENERIC) and the investigation of the approach to a reduced dynamics initiated by Rayleigh, Chapman and Enskog, and Grad. Both investigations have the same structure and use the same type of quantities, which means that the dissipation potential is the rate entropy, and the Rayleighian in Onsager's variational principle becomes the rate thermodynamic potential in the rate GENERIC (dynamical extension of the Onsager principle).

The rate GENERIC can also be seen as GENERIC extended to address preceding stages in the time evolution of macroscopic systems. In the absence of external forces and internal constraints, GENERIC addresses the approach to equilibrium. The rate GENERIC addresses the preceding stage of the time evolution whose outcome, in the absence of external forces,  are the forces driving the approach towards the equilibrium. The rate time evolution is of primary importance not only because it precedes the approach to equilibrium, but also because it does not have to be followed by it.  The rate thermodynamics resulting from the rate time evolution is  applicable also in the presence of external forces or internal constraints that prevent approach to equilibrium.

Illustrations in the field of chemical kinetics and kinetic theory are worked out in Section \ref{illus}. In both theories, we formulate Onsager's variational principle. In particular, we derive a rate Boltzmann equation whose solutions approach the time evolution governed by the Boltzmann equation.

\section{GENERIC}\label{sectiongeneric}

Let us start with the multiscale character of the GENERIC framework. How can we identify a structure that is  common to well established  (for instance, well tested with experimental observations)  mesoscopic dynamical theories? Since a common structure of time evolution equations implies common properties of their solutions, we look first for properties of solutions shared by  well established  mesoscopic time evolution equations.
We focus on the multiscale nature of the time evolution. By this property we mean that by changing the focus of observations from local in time and space to  more global in time and space behavior, results of the coarser observations are 
well described by   an autonomous
theory    that ignores  the  details that are not seen in the coarser observations and manifestly   displays  new emerging overall features that are not seen in the original theory.
The autonomous theory obtained in this way is called  a reduced theory. The reduction involves both a loss and a gain. The details are lost and the overall features are gained.

In particular, externally unforced macroscopic systems reach (as $t\rightarrow\infty$) equilibrium states, where their behavior is well described by the classical equilibrium thermodynamics \cite{callen}. Experimentally observed behavior of externally unforced and internally unconstrained  macroscopic systems can be thus described on two levels: the microscopic level on which the macroscopic systems are seen  as being composed of $\sim 10^{23}$ particles whose time evolution is governed by classical mechanics, and the macroscopic level on which the same physical systems are seen though the eyes of the classical equilibrium thermodynamics. Both levels are autonomous, both have been developed independently on the basis their  own family of experimental observations, and both can be also applied independently. In addition, the experimentally observed behavior of  most macroscopic systems can also be described by a sequence of intermediate (mesoscopic) autonomous levels, as for instance the level of kinetic theory and the level of hydrodynamics \cite{chapman,gk}.

How shall we identify a structure expressing the multiscale nature of the time evolution? There are two paths to take. The first is to recognize  common features in existing  time evolution  equations on well established levels. The second is to investigate in detail solutions of governing equations on a chosen well established single mesoscopic level. On the second path, the first step is to solve completely the governing equations and create the phase portrait (collection of all trajectories). The second step is to recognize in it a pattern that is then interpreted  as the phase portrait corresponding to  the reduced time evolution. Following this path, the  structure  of the time evolution equations that we search is the structure that guarantees the emergence of  patterns in their phase portraits. The obvious difficulties that we meet on  the second path makes  us to turn to the first path on which  governing equations on several well established levels are known from independent investigations. We look for  their  common features.
It was the second path that led to the GENERIC equation discussed in this paper. The GENERIC equation expresses  mathematically the multiscale nature of dynamical theories of macroscopic systems.

The GENERIC equation has three foundation stones: (i) mechanics expressed mathematically in terms of  abstract Hamiltonian system (\cite{clebsch}, \cite{arnold}, (ii) thermodynamics expressed in terms of the gradient dynamics \cite{ch}, \cite{landau-ginzburg}, (iii) the Boltzmann equation \cite{Boltzmann-vorlesungen} providing an example of a well established mesoscopic time evolution in which mechanics and thermodynamics are combined.
An abstract  GENERIC time evolution equation combining the abstract Hamiltonian dynamics with the gradient dynamics appeared first in \cite{dv}. The Boltzmann kinetic equation and the governing equations of hydrodynamics have been put into the form displaying explicitly the GENERIC structure in \cite{grcontmath}, \cite{mor}, \cite{kauf}, and \cite{grpd}.  The name GENERIC (an acronym for General Equation for Non-Equilibrium Reversible-Irreversible Coupling) has been introduced in \cite{go}, \cite{og}. Another name for the same type of structure, introduced in \cite{morrison}, is  a metriplectic structure (referring to a combination of symplectic and metric structures). A difference between GENERIC and metriplectic is that the former uses dissipation potentials while the latter sticks to the dissipative brackets \cite{miroslav-guide}.
The contact geometry has been brought into the GENERIC time evolution  in
\cite{GrmUdeM}, \cite{pkg}, \cite{OMM-I} and stochastic analysis (in particular the large deviation theory in connection with the dissipation potentials) in \cite{mielke-potential}, \cite{mielke-peletier}.

Now we proceed to formulate the GENERIC equation as an abstraction of the Boltzmann equation.
The state variables are denoted by $x$ and $M$
denotes their state space, $x\in M$.  In the particular case of the Boltzmann equation, the state variable is $x=f(\rr,\pp)$, where $\rr$ is a position vector and $\pp$ momentum of one particle and $f(\rr,\pp)$ is the one particle distribution function, and $M$ is the space of all one-particle distribution functions.

In order to derive the equation governing the time evolution of $f(\rr,\vv)$, we begin with the
 Hamiltonian time evolution
\begin{equation}\label{Hamrv}
\left(\begin{array}{cc}\dot{\rr}\\\dot{\pp}\end{array}\right)=\left(\begin{array}{cc}0&1\\-1&0\end{array}\right)
\left(\begin{array}{cc}E_{\rr}\\E_{\pp}\end{array}\right)
\end{equation}
of $(\rr,\pp)$,  where $E(\rr,\pp)$ is the energy of one particle. We use hereafter a shorthand notation $E_{\rr}=\frac{\partial E}{\partial \rr};  E_{\pp}=\frac{\partial E}{\partial \pp}$.
If the macroscopic system under investigation is  a dilute gas, the energy is only the kinetic energy $E(\rr,\pp)=\frac{\pp^2}{2m}$, ($m$ being the mass of one particle) and (\ref{Hamrv}) turns into   $\dot{\rr}=\pp/m=\vv; \dot{\pp}=0$ governing the time evolution of a free particle.

Using the geometrical viewpoint of the Hamiltonian dynamics recalled in \cite{OMM-I,momentum-Euler}, we now  lift (\ref{Hamrv}) to
\begin{equation}\label{Hamf}
\frac{\partial f(\rr,\pp)}{\partial t}=LE_{f(\rr,\pp)},
\end{equation}
governing the time evolution of $f(\rr,\pp)$. The Poisson bivector $L$ is given by the Poisson bracket
\begin{eqnarray}\label{PBf}
\{A,B\}&=&\int d\rr\int d\pp \langle A_f,LB_f \rangle  \nonumber \\
&=&\int d\rr\int d\pp f(\rr,\pp)\left(\frac{\partial A_{f(\rr,\pp)}}{\partial \rr}\frac{\partial B_{f(\rr,\pp)}}{\partial \pp}-\frac{\partial B_{f(\rr,\pp)}}{\partial \rr}\frac{\partial A_{f(\rr,\pp)}}{\partial \pp}\right).\nonumber \\
\end{eqnarray}
With this Poisson bivector $L$, the time evolution equation (\ref{Hamf}) becomes
\begin{equation}\label{Hamff}
\left(\frac{\partial f(\rr,\pp)}{\partial t}\right)_{(Ham)}=-\frac{\partial (fE_{\pp})}{\partial \rr}+\frac{\partial (fE_{\rr})}{\partial \pp}
\end{equation}
which, with the energy $E(\rr,\pp)=\int d\rr\int d\pp f\frac{\pp^2}{2m}$, turns into $\frac{\partial f}{\partial t}=-\frac{\partial (f\vv)}{\partial\rr}$.

In order to explain the reason why  we are making
the lift from $(\rr,\pp)$ to $f(\rr,\pp)$, we recall that we look for  a structure of the time evolution equation which manifestly displays patterns in the phase portrait. We expect that the patterns will be easier to recognize in the trajectories of  particle distribution functions than in the particle trajectories.

The abstract formulation
\begin{equation}\label{GHam}
(\dot{x})_{(Ham)}=\{x,E\}=LE_x
\end{equation}
of the free flow part of the Boltzmann equation  (\ref{Hamf}) is the Hamiltonian part of the abstract GENERIC equation. The operator $L(x)$ in (\ref{GHam}) is called Poisson bivector (see \cite{OMM-I}). The bracket
$\{A,B\}=\langle A_x,L(x)B_x\rangle$ is a Poisson bracket ($A$ and $B$ are real-valued sufficiently regular functionals $M\rightarrow\mathbb{R}$ and $\langle\bullet,\bullet\rangle$ is a pairing in $M$)
satisfying  the following properties:
\begin{subequations}\label{propPB}
\begin{eqnarray}
&&(i)\, \text{skew-symmetry}: \{A,B\}=-\{B,A\} \\
&&(ii)\, \text{bilinarity}: \{rA+sB,C\}=r\{A,C\}+s\{B,C\}, \,\, \text{where} \,\,r\in\mathbb{R}, s\in\mathbb{R}\\
&&(iii)\, \text{Leibnitz \,\,identity}: \,\,\{A,BC\}=\{A,B\}C+\{A,C\}B\\
&&(iv)\, \text{Jacobi\,\, identity}: \,\,\{A,\{B,C\}\}+\{B,\{C,A\}\}+\{C,\{A,B\}\}=0.
\end{eqnarray}
\end{subequations}

Now we turn to the part of the mesoscopic time evolution expressing the lack of details needed in microscopic mechanics. We follow Boltzmann's analysis of the time evolution of dilute gases.
The particles in the dilute gas move freely until they collide with another particle or particles. In order to express collisions in (\ref{Hamrv}), we would have to consider not one but at least two particles (i.e. $(\rr,\pp)$ would have to extend to $(\rr_1,\pp_1,\rr_2,\pp_2)$) and the energy would have to include, in addition to the kinetic energy, also an interaction energy expressing a short range hard core repulsion among the particles. Instead of making this type of extension, we follow Boltzmann and make an extension that already includes elements of the  pattern recognition process. Collisions are events in the time evolution that we expect to be  most consequential for the overall look of the phase portrait. Moreover, details of the trajectories of the two colliding particles are not expected to influence the pattern in the phase portrait. The two particle collisions can be taken into account in the time evolution of $f(\rr,\pp)$ as a pointwise gain-loss balance of momenta that preserves the total momentum and the total kinetic energy. These considerations (see more details in Section \ref{RTT}) lead to
\begin{equation}\label{GdissB}
\left(\frac{\partial f}{\partial t}\right)_{(diss)}= \Xi_{f^*(\rr,\pp)}|_{f^*(\rr,\pp)=S_{f(\rr,\pp)}}
\end{equation}
where $\Xi(f,f^*)$ is a dissipation potential and $S(f)=-k_B\int d\rr\int d\pp f(\rr,\pp)\ln f(\rr,\pp)$ is the Boltzmann entropy; $k_B$ is the Boltzmann constant. The explicit form of the dissipation potential $\Xi(f,f^*)$ is given in Section \ref{RTT}.

We now list the properties of dissipation potential $\Xi$. The dissipation potential $\Xi:T^*M\rightarrow \mathbb{R}$ usually depends on $x^*$ only through its dependence on $\mathcal{K}x^*$, where $\mathcal{K}$ is a linear operator\footnote{This is similar to a decomposition of dissipative brackets \cite{hco}.}.
For example in fluid mechanics, $\mathcal{K}$ is the  gradient $\frac{\partial}{\partial\rr}$, and in chemical kinetics (see Section \ref{sectionchemkin}), it is the stoichiometric matrix. We shall use the notation $X^*=\mathcal{K}x^*$ and call it, in accordance with an established terminology in nonequilibrium thermodynamics,  a thermodynamic force. Moreover, dissipation potentials are required to satisfy:
\begin{eqnarray}\label{Xiprop}
&&(i)\, \langle x^*,\Xi_{x^*}\rangle=a\langle X^*,\Xi_{X^*}\rangle,\,\,\text{where} \,\,a\in\mathbb{R}^{+}\nonumber \\
&&(ii)\, \Xi(x,X^*)|_{X^*=0}=0;\text{\,\,for\,\,all}\,\,x\in M\nonumber \\
&&(iii)\, \Xi\,\,\text{reaches \,\,its\,\,minimum\,\,at}\,\,X^*=0\nonumber \\
&&(iv)\, \Xi(x,X^*)\,\,\text{is \,\,a\,\,convex\,\,function\,\,of}\,\,X^*\,\,\text{in\,\,a\,\,neighborhood\,\,of}\,\,X^*=0,\,\forall x\in M.
\end{eqnarray}

The abstract formulation
\begin{equation}\label{Gdiss}
(\dot{x})_{(diss)}=\Xi_{x^*}(x,X^*(x^*))|_{x^*=S_x}
\end{equation}
of (\ref{GdissB}) is taken to be the  dissipative part of the abstract GENERIC equation.
We note that for $X^*$ that are close to zero all dissipation potentials become
quadratic dissipation potentials $\Xi(x^*)=\frac{1}{2}\langle X^*,\Lambda(x) X^*\rangle$, where $\Lambda(x)$ is a positive definite operator, and (\ref{Gdiss}) becomes $(\dot{x})^{(diss)}=(\mathcal{K})^T\Lambda S_x$, which is in the form of a metriplectic system \cite{mor}.

The complete GENERIC equation
\begin{equation}\label{generic}
\dot{x}=(\dot{x})_{(Ham)}+(\dot{x})_{(diss)}=L(x)E_x+\left[\Xi_{x^*}(x,X^*)\right]_{x^*=S_x}
\end{equation}
combines the Hamiltonian part (\ref{GHam}) and the dissipative part (\ref{Gdiss}). The combination brings  new requirements on the Poisson bivector $L$ and the dissipation potential $\Xi$. Both these quantities are required to be complementary  degenerate in the sense that
\begin{eqnarray}\label{degg}
&&LS_x=0\,\,\forall\,\, x\in M\nonumber \\
&&\langle E_x,\left[\Xi_{x^*}(x,X^*)\right]_{x^*=S_x}\rangle=0\,\,\text{and}\,\, \left[\Xi_{x^*}(x,X^*)\right]_{x^*=E_x}=0,\,\,\forall\,\, x\in M.\nonumber \\
\end{eqnarray}
Equation (\ref{generic}) in which $x$, $E(x)$, $S(x)$, $\{A,B\}$, and $\Xi(x,X^*)$ are specified is called a particular realization of the GENERIC equation (\ref{generic}).

Using the standard terminology, the entropy $S(x)$ is required to be a Casimir and the energy $E(x)$ a dissipation Casimir.
Casimirs are functionals $C(x)$ for which $LC_x=0,\,\,\forall x\in M$, whereas functionals $ \mathbb{C}(x)$ for which
 $\langle \mathbb{C}_x,\left[\Xi_{x^*}(x,X^*)\right]_{x^*=\mathbb{S}_x}\rangle=0$ and
$\left[\Xi_{x^*}(x,X^*)\right]_{x^*=\mathbb{C}_x}=0\,\,\forall\,\, x\in M$ are called dissipation Casimirs. In the particular case of the quadratic dissipation potential $\Xi(x,X^*)=\frac{1}{2}\langle X^*,\Lambda(x) X^* \rangle$,  the  requirement of the dissipation degeneracy takes the form $\Lambda E_x=0$ for all $x\in M$.
The degeneracy requirement (\ref{degg}) is needed in order that the pattern in the phase portrait established as $t\rightarrow\infty$ is  the pattern of equilibrium thermodynamics (see Section \ref{et}).

The physical content of GENERIC is expressed in two geometrical structures (symplectic and gradient) and three potentials, entropy, energy, and number of moles (or actually total mass). In the contact geometry formulation  developed in Part I \cite{OMM-I}, the geometrical structures enter in the contact Hamiltonian and the potentials in the Legendre manifold on which the GENERIC time evolution takes place.

\subsection{Approach to equilibrium, MaxEnt}\label{sectionmaxent}
Let us now focus on how GENERIC equation (\ref{generic}) approaches the equilibrium and how it corresponds with the principle of maximum entropy (MaxEnt). Our goal is to recognize patterns in the phase portraits generated by (\ref{generic}). The simplest and the most obvious start is to look for fixed points reached as $t\rightarrow \infty$. The pattern formed by asymptotically reached fixed points will be  the first pattern in the phase portrait that we shall interpret physically as a pattern representing a reduced theory.

Being inspired by Boltzmann's analysis of solutions of the Boltzmann equation, we look for a Lyapunov function that will lead us to fixed points. Already the way the GENERIC equation (\ref{generic}) is written suggests to consider  the entropy $S(x)$ as  the first obvious candidate for the Lyapunov function. We shall now explore consequences of this choice. Physical meaning and consequences of another candidate for the Lyapunov function will be investigated in Section \ref{sectionRgeneric}.

 We recall that in the original Boltzmann analysis of solutions to  the Boltzmann equation the collision (dissipative) part did not have the form (\ref{GdissB}) but a form (called a Boltzmann collision term, see (\ref{kinB})), which arises in an explicit analysis of  mechanics of collisions. Boltzmann's entropy $S(f)=-k_B\int d\rr\int d\pp f \ln f$ arises  in Boltzmann's H-theorem, which essentially consists of recasting the Boltzmann collision term into the form of Equation (\ref{GdissB}). This is because Equation (\ref{GdissB}) together with the degeneracy requirement (\ref{degg}) implies that
\begin{equation}\label{Gpropp}
\dot{S}=\int d\rr\int d\pp S_f\left[\Xi_{f^*}(f,f^*)\right]_{f^*=S_f}\geq 0.
\end{equation}
The inequality is a direct consequence of the properties of the dissipation potential (\ref{Xiprop}).

Moreover, the second law of thermodynamics, (\ref{Gpropp}), holds also for solutions of the abstract GENERIC equation (\ref{generic})
\begin{subequations}\label{GGprop}
\begin{eqnarray}
\dot{S}&=&\langle S_x,[\Xi_{x^*}(x,X^*)]_{x^*=S_x}\rangle=a\langle X^*,\Xi_{X^*}\rangle|_{x^*=S_x}\geq 0 \\
\dot{E}&=&0.
\end{eqnarray}
\end{subequations}
For the later use, we also rewrite the entropy inequality into the form
\begin{equation}\label{GGpp}
\dot{S}=\langle S_x,[\Xi_{x^*}(x,x^*)]_{x^*=S_x}\rangle =\left[\Xi(x,x^*)+\Xi^{\dag}(x,x^{\dag})|_{x^{\dag}=\Xi_{x^*}(x,x^*)}\right]_{x^*=S_x}
\end{equation}
where $\Xi^{\dag}(x,x^{\dag})$ is the Legendre transformation of $\Xi(x,x^*)$ in the variable $x^*$. The second equality in (\ref{GGpp}) follows directly from the definition of the Legendre transformation. Actually, the Fenchel equality reads
\begin{equation}\label{fenchel}
 \langle x^*,x^\dagger \rangle \leq \Xi(x,x^*)+\Xi^{\dag}(x,x^{\dag}),
\end{equation}
and the equality holds when $x^\dagger$ and $x^*$ are related via the Legendre transform.

The relations  (\ref{GGprop})  together with the requirements of the concavity of the entropy $S(x)$ and  convexity of the energy $E(x)$ makes the thermodynamic potential
\begin{equation}\label{Phi}
\Phi(x; e^*,n^*)=-S(x)+e^*E(x)+n^*N(x)
\end{equation}
a Lyapunov function for the approach to equilibrium states $\widehat{x}(e^*,n^*)$ that are solutions to $\Phi_x=0$ and form a manifold
\begin{equation}\label{eqManifold}
\widehat{\mathcal{M}}=\{x\in M|\Phi_x=0\},
\end{equation}
called the equilibrium manifold.\footnote{Potential $\Phi$ is a Lyapunov function only when the system under consideration is isolated. For open systems, it has to be generalized as in \cite{vitek-entropy}.} The function $N(x)$ appearing in (\ref{Phi}) is a real-valued and sufficiently regular function (or a collection of functions) that is (are) both Casimirs and dissipation Casimirs \cite{OMM-I}, and thus remain unchanged during the time evolution governed by (\ref{generic}). In the context of mesoscopic dynamics,  one  example of such function $N(x)$ is the total number of moles (in kinetic theory, $N(f)=\int d\rr\int d\pp f(\rr,\pp)$). The quantities $e^*$ and $n^*$ are real numbers playing the role of Lagrange multipliers in the minimization of the thermodynamic potential $\Phi(x;e^*,n^*)$. Their physical meaning will be discussed below in Section \ref{et}.
The complementary degeneracy (\ref{degg}) of the Poisson and gradient structures allows us to write (\ref{generic}) in the form
\begin{equation}\label{Generic}
\dot{x}=\frac{1}{e^*}L\Phi_x-\left[\Xi_{x^*}(x,x^*)\right]_{x^*=\Phi_x}
\end{equation}
which displays the thermodynamic potential (\ref{Phi}) as the Lyapunov function.

A closer examination of trajectories in the  neighborhood of    the equilibrium states $\widehat{x}(e^*,n^*)$ reveals that the time inequality (\ref{GGprop}) does not suffice to prove the approach to  $\widehat{x}(e^*,n^*)$.
We illustrate it on the example of the Boltzmann equation (i.e. $x$ is a one particle distribution function).
We see easily  that the trajectories approaching the equilibrium states $\widehat{x}(e^*,n^*)$
have to avoid the manifold
\begin{equation}\label{disseq}
\mathcal{M}^{(diss)}=\{x\in M|X^*=0\},
\end{equation}
called the dissipation-equilibrium manifold.  This is because $\mathcal{M}^{(diss)}$ is an invariant manifold and because
$X^*(f)=0$  for all $f\in \mathcal{M}^{(diss)}$  implies that there is  no dissipation force inside  $\mathcal{M}^{(diss)}$ that would drive the approach of $f\in\mathcal{M}^{(diss)}$ to
$\widehat{x}(e^*,n^*)\in\widehat{\mathcal{M}}\subset\mathcal{M}^{(diss)}$.

In order to see the invariance of  $\mathcal{M}^{(diss)}$, we recall (see more in Section \ref{illus})  that
in the case of the Boltzmann equation, the dissipation equilibrium manifold $\mathcal{M}^{(diss)}$  is composed of local Maxwellian distribution functions and the thermodynamic equilibrium manifold $\widehat{M}$ is composed of total Maxwellian distribution functions. The Maxwell distribution $f^{(Maxw)}(\rr,\pp)$ is defined as a distribution function whose conjugate  $f^{(Maxw)^*}(\rr,\pp)=S_{f^{(Maxw)}(\rr,\pp)} =a+\langle\bb,\pp
\rangle+c\pp^2$, where $S(f)$ is the Boltzmann entropy $S(f)=-k_B\int d\rr\int d\pp f\ln f$ and $a,\bb,c$ are parameters. The Maxwell distribution is called  local if $(a,\bb,c)$ are functions of $\rr$ such that $\int d\pp f^{(Maxw)}=\int d\pp f; \int d\pp f^{(Maxw)}\pp=\int d\pp f \pp;   \int d\pp f^{(Maxw)}\pp^2=\int d\vv f \pp^2$. The Maxwell distribution function is called total if  $(a,\bb,c)$ are constants such that $\bb=0$ and  $\int d\pp \int d\rr f^{(Maxw)}=\int d\pp \int d\rr f;   \int d\pp \int d\rr f^{(Maxw)}\pp^2=\int d\pp \int d\rr f \pp^2$. In addition, we note that
solutions to $\frac{\partial f(\rr,\pp)}{\partial t}=LE_{f(\rr,\pp)}$ are $f_0(\rr-\pp t,\pp)$, where $f_0(\rr,\pp)=f(\rr,\pp,t)|_{t=0}$).

The presence of the Hamiltonian part of the vector field in combination with the dissipation part $\Xi_{x^*}$ prevents the trajectories $x(t)$ to enter into $\mathcal{M}^{(diss)}$, except at the final destination
$\widehat{x}(e^*,n^*)\in \widehat{\mathcal{M}}\subset \mathcal{M}^{(diss)}$. The Hamiltonian part of the vector field does not generate any dissipation by itself, but in combination with the gradient part of the vector field, it enhances it. This role of the Hamiltonian dynamics in the dissipative time evolution has been proven for the Boltzmann equation in \cite{grad-h}, \cite{desvilettes-villani}. It is likely that this dissipation mechanism (Grad-Desvillettes-Villani enhancement of dissipation) plays a key role in the emergence of the time irreversibility in the microscopic dynamics. A very small microscopic  instability (a microturbulence) arising in microscopic dynamics is gradually enhanced by the Grad-Desvillettes-Villani mechanism into the dissipation appearing in the GENERIC equation (\ref{generic}), that then drives the approach to equilibrium states $\widehat{x}(e^*,n^*)$. This  conjecture is at present supported only by results  about solutions to  the Boltzmann equation in which the gradient part of the vector field generating the dissipation is already present but is not strong enough by itself to drive solutions to the equilibrium states. Only the vector field that combines the weak dissipative part with the non-dissipative Hamiltonian part brings solutions to the Boltzmann equation to the equilibrium states (to the total Maxwellian distribution functions) \cite{desvilettes-villani}.

\subsection{Equilibrium thermodynamics}\label{et}

We have shown that the time evolution generated by (\ref{generic}) terminates at equilibrium states $\widehat{x}(e^*,n^*)\in \widehat{\mathcal{M}}\subset M$ (see (\ref{eqManifold})).
The manifold  $\widehat{\mathcal{M}}$ is the pattern that we have recognized in the phase portrait generated by (\ref{generic}). We can arrive at $\widehat{\mathcal{M}}$ either by following the GENERIC time evolution or alternatively simply by solving
\begin{equation}\label{maxent}
\Phi_x(x,e^*,n^*)=0,
\end{equation}
i.e. by maximizing the entropy $S(x)$ subjected to constraints $E(x)$ and $N(x)$. The latter route to $\widehat{\mathcal{M}}$ is called Maximum Entropy principle or briefly \textit{MaxEnt principle} \cite{jaynes}.

It may  seem  strange that a pattern, that is something endowed with an order, emerges in the process of maximizing entropy that increases disorder. Here we see the importance of constraints in the maximization. Allegorically speaking, the constraints are the bricks from which the pattern is built. The rest is a sand. The bricks  are assembled into a pattern in  the maximization process. The entropy maximization makes the details (i.e. the sand) maximally irrelevant so that they can be eventually (when the maximization process is completed) eliminated.

The two potentials $E(x)$ and $N(x)$ representing the constraint as well as the entropy $S(x)$ have roots in the time evolution. The entropy $S(x)$ drives the approach to $\widehat{\mathcal{M}}$, the energy $E(x)$ and the number of moles $N(x)$ are constants of motion. When we do not know the time evolution and use only MaxEnt  to pass from the state space $M$ to $\widehat{\mathcal{M}}$, then all three potential have to be postulated.
A notable example of using only the MaxEnt principle to reach $\widehat{\mathcal{M}}$ is
the Gibbs equilibrium statistical mechanics in which $M$ is the completely microscopic space with n-particle distribution functions, $n\sim 10^{23}$ as its elements. The time evolution enters only indirectly in the postulated Gibbs entropy, that is assumed to be universally applicable to all macroscopic systems and that expresses  indirectly an  assumed  ergodicity of the phase portrait.

Now we ask the question of what is the autonomous reduced level that the pattern recognized in the equilibrium states $\widehat{\mathcal{M}}$ represents.  The quantities $e^*$ and $n^*$,  that serve in MaxEnt as Lagrange multipliers, parametrize the pattern and thus  serve  as state variables in the reduced theory.   The structure on $\widehat{\mathcal{M}}$ that is inherited from GENERIC (\ref{generic}) is  thus:

(i) no time evolution,

(ii) three state variables
  \begin{equation}\label{eqstvar}
(S^*,e^*,n^*)
\end{equation}
where $S^*\in\mathbb{R}, e^*\in\mathbb{R}, n^*\in\mathbb{R}$, and

(iii) a relation, called a fundamental thermodynamic relation,  among them
\begin{equation}\label{eqrelat}
S^*=S^*(e^*,n^*)
\end{equation}
where $S^*(e^*,n^*)= \Phi(\widehat{x}(e^*,n^*);e^*,n^*)$.

The relation (\ref{eqrelat}) does   not however represent yet the complete equilibrium thermodynamics. What is missing is the physical interpretation of (\ref{eqstvar})? Macroscopic systems at equilibrium interact with their environment by interchanging heat and/or mechanical macroscopic work. The interactions are made on the boundaries of the systems. In the investigation of dynamics that led us to (\ref{generic}), we have ignored the boundaries by considering the macroscopic systems to be either infinite or finite with periodic boundary conditions. We shall not begin to discuss mesoscopic and microscopic dynamics with boundaries, but we shall  include boundaries directly into (\ref{eqstvar}) and (\ref{eqrelat}). We take the boundaries into account in  the following three steps:

\begin{enumerate}
\item
We transform (\ref{eqrelat}) into
\begin{equation}\label{eq1}
S=S(E,N)
\end{equation}
by the Legendre transformation \\(i.e. $S(E,N)=[-S^*(e^*,n^*)+e^*E+n^*N]_{(e^*,n^*)=\widehat{(e^*,n^*)}}$, where $\widehat{(e^*,n^*)}$ is a solution to $(-S^*(e^*,n^*)+e^*E+n^*N)_{e^*}=0$ and $(-S^*(e^*,n^*)+e^*E+n^*N)_{n^*}=0$).
Next, we extend (\ref{eq1}) into
\begin{equation}\label{eq2}
S=S(E,N,V)
\end{equation}
where $V$ is a new state variable representing the volume of the macroscopic system under consideration. Until now, the  entropy $S$, the energy $E$, and the number of moles $N$ have been their  values per unit volume.

\item
The state variables $(S,E,N)$ are extensive. This means that if  we make a transformation  $V\rightarrow \lambda V$, where $\lambda \in \mathbb{R}$,  then also $E\rightarrow \lambda E; N\rightarrow \lambda N$, and $S\rightarrow \lambda S$.  The extensivity, if applied to the relation (\ref{eq2}), implies that the function $S(E,N,V)$ is a 1-homogeneous function, i.e. $\lambda S=S(\lambda E, \lambda N, \lambda V)$.  The 1-homogeneity then implies
\begin{equation}\label{eq3}
S=e^*E+n^*N+v^*V,
\end{equation}
which in turn  implies
\begin{equation}\label{eq4}
S^*=v^*.
\end{equation}

\item
Finally, we assume that the walls that separate   macroscopic systems  and that   either freely pass or complete stop passing  $E, N$, and $V$ are readily available.  The passage of the internal energy is the passage of heat, the passage of $N$ is made by using appropriate membranes, and the volume is changed, for instance, by moving  a piston in a cylinder.

\end{enumerate}

With the above three steps, with which we have extended (\ref{eqrelat}), implied by GENERIC, we have arrived at a formulation of the complete classical equilibrium thermodynamics.  Using the standard terminology, we have that $(e^*,n^*,v^*)=(\frac{1}{T},-\frac{\mu}{T}, -\frac{P}{T})$, where $T$ is the absolute temperature, $\mu$ is the chemical potential, and $P$ is the pressure.

The state variables (\ref{eqstvar}) are directly measurable.
Let us recall for example the measurement of the temperature $T$. A thermometer is a macroscopic system for which the fundamental thermodynamic relation  is known (from  experimental observations of the relation among, for instance $P,V$, and $T$), and which is surrounded by  walls that  prevent changes in $V$ and $N$ while allowing passage of internal energy $E$.  If we now put the thermometer into  contact with a system under investigation and surround both systems with a wall that does not pass $E$, then MaxEnt implies that the temperatures  inside and outside of the thermometer are the same. Since we know the fundamental thermodynamic relation  inside the thermometer, we can read the temperature inside the thermometer in, for example,  the pressure or the volume measured inside  the thermometer.

Summing up, the classical equilibrium thermodynamics is a theory reduced from the GENERIC equation (\ref{Generic}) (by focusing on the pattern in its phase portrait composed of fixed points) together  with an extension that   addresses  the  size and the  boundaries of the macroscopic systems (aspects that are ignored  in GENERIC). In the following section, we extend GENERIC similarly as Extended Irreversible Thermodynamics \cite{jou-eit} extends Classical Irreversible Thermodynamics \cite{dgm}.

\section{Rate GENERIC}\label{sectionRgeneric}

In this section,
we are aiming at the stages of the time evolution that precede the approach to equilibrium. In other words, we are aiming at the stages where the forces described by GENERIC are created. We anticipate to find such time evolution, called a rate GENERIC time evolution, as the GENERIC time evolution lifted to the iterated tangent and cotangent bundles (route 1), or as the GENERIC  time evolution  describing the approach to equilibrium  on  levels involving more details (route 2). We have started to follow the first route in Part I \cite{OMM-I} and continue in Section \ref{TD0}. Iterated tangent and cotangent bundles provide indeed a stage  for bringing  a new physics addressing the time evolution of forces. On the second route, that we follow in Section \ref{TD},  we begin with the GENERIC  equation formulated on a more microscopic level. A new (more microscopic) physics is thus entering this route at the outset of the investigation. The more microscopic GENERIC equation is subsequently recast into a  hierarchy (like for example recasting the Boltzmann equation into the Grad hierarchy - see Section \ref{illus}). The rate GENERIC time evolution then closes the hierarchy.

The
rate GENERIC provides also  a new look at the pattern recognition process in reductions. In comparison with the previous section, where the reduced (target) level is the level of equilibrium thermodynamics, on which no time evolution takes place, in this section we investigate reductions to levels on which a reduced (more macroscopic) time evolution does take place. This generalization  considerably enlarges the applicability. In particular, macroscopic systems subjected to external forces or internal constraints can also be considered. While such systems are prevented from reaching equilibrium states and thus equilibrium thermodynamics is not applicable to them, they, in general,  experience reductions to more macroscopic levels. For instance, a horizontal layer of a fluid that is heated from below (Rayleigh-B\'{e}nard system) can be well described on the level of hydrodynamics (and, of course, on more microscopic levels as e.g. the completely microscopic level) \cite{lebon-understanding}. The pattern recognition process itself can be made either in the phase portrait composed of trajectories in the state space  or in the phase portrait composed of trajectories of vector fields. In the former pattern recognition process, we seek invariant manifolds in the state space  on which a slower time evolution takes place.
In the latter pattern recognition process, we search for fixed points (that are the vector fields generating the reduced time evolution).
The first viewpoint of reductions follows the pioneering works of Chapman and Enskog \cite{chapman-cowling} in their investigation of the reduction of the Boltzmann kinetic equation to hydrodynamics (see also \cite{gk}). The second viewpoint of reductions follows Grad \cite{Gradhier} in his investigation (that begins with a hierarchy reformulation of the Boltzmann kinetic equation) of the same physical problem. We shall demonstrate that this second viewpoint  can  also be seen as an extension of  Onsager's variational principle.
In this paper, we concentrate searching for fixed points in the phase portrait of vector fields. The Chapman and Enskog approach is only briefly recalled in Section \ref{TD}.

Regarding the notation, we keep $M$ to denote the state space, $x\in M$. Moreover, we introduce a new (more macroscopic) space  $N$  with elements $y\in M$. Both spaces $M$ and $N$ are assumed to be linear spaces, $M^*$ and $N^*$ are their duals, $x^*\in M^*$, $y^*\in N^*$.
In order to simplify our terminology,  we shall call the level on which $x$ serves as the state variable an $M$-level, similarly, we shall use an $N$-level  and an $N^{(eth)}$-level, where $N^{(eth)}$ is the state space of the equilibrium thermodynamics with elements $(E,N)\in N^{(eth)}$.

\subsection{Geometric lifts}\label{TD0}
We shall arrive at the rate GENERIC time evolution equation in two steps.
First, we note that the force  playing the most important role in GENERIC is the gradient of entropy, $S_x(x)$. This is the force that drives the reduction to equilibrium. Its importance  is then expressed in the geometrical formulation by regarding $S_x(x)$ as a conjugate of $x$. Consequently, we see that if we are interested in reaching beyond the approach to equilibrium to an approach to forces governing it, we have to turn our interest to the time evolution of $x^*$.
By lifting  GENERIC to the cotangent bundle \cite{OMM-I}, we obtain
 \begin{equation}\label{rgeneric}
\dot{x^*}=Hess^{*-1}(x^*)\left[\left(\langle x^*,L(x)E_x(x)\rangle +\Xi(x,x^*)\right)_{x^*}\right]_{x=S^*_{x^*}(x^*)}
\end{equation}
that is an equation governing the time evolution of $x^*$;  $Hess^*(x^*)$ is the Hessian  of $S^*(x^*)$ that is the  Legendre transformation of $S(x)$, and $Hess^{*-1}(x^*)$ is its inverse.
We emphasize that (\ref{rgeneric}) is completely equivalent to (\ref{generic}) when restricted to $x^*=S_x$. However, if this is not the case, if $x^*$ is considered as an independent variable, then (\ref{rgeneric}) is a new equation.

In the second step, we complete the formulation of the rate GENERIC by leaving the setting of GENERIC that approaches equilibrium while focusing on the time evolution driven by  $\Xi_{x^*}(x,x^*)$. We regard such time evolution  as an autonomous model of the time evolution (we call it a rate time evolution) that terminates when the driving force  $\Xi_{x^*}(x,x^*)$  disappears. When the rate time evolution is completed, the time evolution of the macroscopic systems under investigation still continues. This next stage, that follows the stage of the rate time evolution, is called a reduced time evolution. In the case when external forces and internal constraints are absent, the reduced time evolution is the GENERIC approach to equilibrium. In the case when  external forces and internal constraints are present, the reduced time evolution is a time evolution that is driven by the external influences.

The simplest time evolution in $M^*$ driven by the gradient $\Xi_{x^*}(x,x^*)$ is
\begin{subequations}\label{rG}
\begin{equation}\label{GENERIC}
\dot{x^*}= \mathbb{G}(x^*)\Psi_{x^*}(x^*,\JJ).
\end{equation}
The time evolution in the reduced state space $N$ is governed by
\begin{equation}\label{GENERIC1}
\dot{y}=Y(x,x^*).
\end{equation}
\end{subequations}
We shall call this pair of equations (\ref{rG}) a \textit{rate GENERIC equation}. The reduced time evolution that follows the rate time evolution governed by (\ref{GENERIC}) is governed by
\begin{equation}\label{GENERIC2}
\dot{y}=Y(x,x^*)|_{x^*=\widehat{x}^*(\JJ)}
\end{equation}
where $\widehat{x}^*(\JJ)$ is a solution to
\begin{equation}\label{minrent}
\Psi_{x^*}(x^*,\JJ)=0.
\end{equation}
This approach has already been used to describe phase inversion \cite{MG-CR} and non-Newtonian fluids with hysteresis in the stress-strain relation \cite{nonconvex}.

We now explain the meaning of the quantities appearing in (\ref{rG}). The operator $\mathbb{G}:M\rightarrow M^*$ is positive definite and serves as a metric tensor that transforms a vector field to a covector field $\dot{x}^*$.
The potential  $\Psi: M^*\times N\rightarrow \mathbb{R}$
\begin{equation}\label{rPsi}
\Psi(x^*,\JJ) =-\mathfrak{S}(x^*)+\langle \XX^*(x^*),\JJ \rangle
\end{equation}
is called a rate thermodynamic potential (compare with the thermodynamic potential  (\ref{Phi})). By $\langle\bullet,\bullet\rangle$ in (\ref{rPsi}) we denote a scalar product in the space $N$ (or duality in distributions). The rate thermodynamic potential $\Psi(x^*,\JJ)$ is a convex function of $x^*$. The rate entropy $\mathfrak{S}(x^*)$ is required  to be a concave function of $x^*$.
The thermodynamic forces $\XX^*:M^*\rightarrow N^*; \XX^*=\mathcal{K}x^*$ are those introduced already in (\ref{Xiprop}), and $\mathcal{K}$ is a linear operator. The Lagrange multipliers  $\JJ \in N$ are fluxes. The flux $Y:M^*\rightarrow TN$ is the vector field generating the reduced time evolution governed by (\ref{GENERIC1}).
The state $\widehat{x}^*(\JJ)\in M^*$ is the state at which the rate time evolution terminates.
We shall see below in Section \ref{OVP}  that Equation (\ref{minrent}), in which the potential $\Psi$ is called Rayleighian, becomes the Onsager's variational  principle. We can therefore regard the rate GENERIC (\ref{GENERIC}), (\ref{GENERIC1}) as a dynamical extension of Onsager' variational principle.

When we compare GENERIC (\ref{generic}) with the rate GENERIC (\ref{rG}), we see first of all that (\ref{generic}) describes the time evolution in $M$ while (\ref{GENERIC}) in $M^*$.
In the thermodynamic potentials
(\ref{rPsi}) and (\ref{Phi}), the entropy $S(x)$ is replaced by the rate entropy $\mathfrak{S}(x^*)$, the constraints (i.e. the energy $E(x)$ and the number of moles $N(x)$ in (\ref{Phi})) by the forces $\XX^*(x^*)$, and the Lagrange multipliers $(e^*,n^*)$  by the fluxes $\JJ$. Moreover, in GENERIC, the reduced time evolution governed by (\ref{GENERIC1}) is no time evolution governed by
\begin{eqnarray}\label{gG1}
\dot{E}&=& \langle E_x,\left[LE(x)_x-\Xi_{x^*}(x,x^*)|_{x^*=S_x}\right]_{x=\widehat{x}(e^*,n^*)}\rangle = 0,\nonumber \\
\dot{N}&=& \langle N_x,\left[LE(x)_x-\Xi_{x^*}(x,x^*)|_{x^*=S_x}\right]_{x=\widehat{x}(e^*,n^*)}\rangle =0.
\end{eqnarray}

The rate GENERIC equation (\ref{rG}) has five building blocks: the operator $\mathbb{G}$, the rate entropy $\mathfrak{S}$, the forces $\XX^*$, the fluxes $\JJ$, and the flux $Y$.
In the rest of this section, we shall explore several ways that can be taken to specify them. At this point we only note that by comparing (\ref{GENERIC}) with (\ref{rgeneric}), we see one particular realization of (\ref{GENERIC}) with
\begin{eqnarray}\label{partrG}
M&\equiv & N\nonumber \\
\Psi(x,x^*)&=& \Xi(x,x^*) +\langle x^*,LE_x\rangle\nonumber \\
\mathbb{G} &=& Hess^{*-1}\nonumber \\
Y&=& LE_x-\Xi_{x^*}(x,x^*)|_{x^*=S_x}.
\end{eqnarray}

Before discussing  the physics involved in (\ref{rG}) in more details,  we investigate some  properties of its solutions.
We note that  the rate GENERIC equation (\ref{GENERIC}) is formally a particular realization of the abstract GENERIC equation (\ref{generic}) in which the Hamiltonian part is missing,  the rate entropy $\mathfrak{S}$ plays the role of the entropy  $S$, the rate thermodynamic  potential $\Psi$ plays the role of the thermodynamic potential $\Phi$, and the dissipation potential is $\frac{1}{2}\langle \Psi_{x^*},\mathbb{G} \Psi_{x^*}\rangle$. Consequently, the rate thermodynamic potential $-\Psi$ plays the role of the Lyapunov function for the approach of solutions of (\ref{GENERIC}) to the rate equilibrium states $\widehat{x^*}(\JJ)$ that are  solutions to (\ref{minrent}).
Indeed, the inequality
\begin{equation}\label{Psidot}
\dot{\Psi}= \langle \Psi_{x^*},\mathbb{G}\Psi_{x^*} \rangle \geq 0
\end{equation}
and the convexity of $-\Psi$
are consequences of  the requirements listed in the text that follows Eq.(\ref{GENERIC}). We note that in GENERIC (\ref{generic}), the convexity of the Lyapunov function is a consequence of the convexity of the thermodynamic potential $\Phi$ (i.e. thermodynamic stability), and the time inequality is a consequence of the convexity of the dissipation potential (i.e. dynamic stability). In the rate GENERIC equation  (\ref{GENERIC}), with the building blocks specified in  (\ref{partrG}), the roles of the thermodynamic and dynamic stabilities are reversed.

The manifold
\begin{equation}\label{manreq}
\widehat{\mathfrak{M}}=\{x^* \in M^*| \Psi_{x^*}(x^*,\JJ)=0\}
\end{equation}
is called a \textit{rate equilibrium manifold}. This manifold can be reached either by following the rate GENERIC time evolution generated by (\ref{GENERIC}) or by minimizing $\Psi$ (i.e. minimizing the rate entropy $\mathfrak{S}(x^*)$) subjected to constraints $\JJ(x^*)$. The minimization of the rate entropy subjected to constraints is called a Minimum Rate entropy principle or briefly \textit{MinRent principle}.

As we have already discussed  at the beginning of Section \ref{sectiongeneric}, the GENERIC equation (\ref{generic}),  and now also the rate GENERIC equations (\ref{rG}) are intended to summarize  experience collected about  certain behavior of macroscopic systems. In the case of (\ref{rG}), it is mainly the experience and the physical insights acquired in hydrodynamics in \cite{Rayleigh} and, as we shall see below, also  in the  analysis of Grad's hierarchy reformulation of the Boltzmann kinetic equation.

\subsection{Onsager's variational principle}\label{OVP}

Macroscopic systems subjected  to external forces  seem to  react in a way that minimizes their resistance \cite{lebon-understanding}. Attempts to  formulate this observation more clearly has led to Onsager's variational principle \cite{onsager1931}. Our goal in this section is to recall this principle, recall (without claiming completeness) its various formulations, and demonstrate that the rate GENERIC (\ref{rG}) is its dynamical extension. Onsager's variational principle is presented as a reduction of a dynamical theory to another dynamical theory involving less details.
As the zero law of thermodynamics (existence of the approach to equilibrium) provides foundation of the classical equilibrium thermodynamics, the zero law of the rate thermodynamics (existence of the approach to slow dynamics involving less details) provides  foundation of the rate thermodynamics (i.e. foundation of Onsager's variational principle). The relation of the rate thermodynamics to Onsager's variational principle makes it also possible to use the   experience collected in its investigation and its many applications in finding the building blocks of (\ref{GENERIC}) representing specific macroscopic systems.

\subsubsection{Purely dissipative systems}\label{sec.GD}
Consider a purely dissipative gradient dynamics of some state variables $x$, $\dot{x}=\Xi_{x^*}|_{x^*=S_x}$. The Legendre-Fenchel transformation of the dissipation potential is
\begin{equation}
\Xi^*(\dot{x}) = \sup_{x^*} (-\Xi(x^*) + x^* \dot{x}),
\end{equation}
and it fulfills that $\Xi^*_{\dot{x}} = x^*$. Note that $x^* \dot{x} = \dot{S}$ can be interpreted as the entropy production. This is a generalization of the original Onsager variational principle to non-quadratic dissipation functions \cite{gyarmati,mielke-potential}.

The inverse transformation gives that
\begin{equation}
\Xi(x^*) = \sup_{\dot{x}} (-\Xi^*(\dot{x}) + x^* \dot{x}),
\end{equation}
which means that $\dot{S}-\Xi^*(\dot{x})$ attains its maximum, or that $\Xi^*(\dot{x})-\dot{S}$ attains its minimum. This is a generalization of Onsager's principle of least dissipation \cite{onsager1931}. Onsager's variational principle and his principle of least dissipation can be thus seen as consequences of gradient dynamics (the irreversible part of GENERIC).

Onsager's variational principle can be also seen from a geometrical point of view. By minimization of the action
\begin{equation}
\int_{t_0}^{t_1} \Xi^*(x,x^*) dt,
\end{equation}
we obtain the Hamilton equations
\begin{equation}
\dot{x} = \Xi_{x^*}
\qquad \mbox{and} \qquad
\dot{x^*} = -\Xi_{x},
\end{equation}
which are accompanied with the stationary Hamilton-Jacobi equation $\Xi(x,x^*)=const$. Vice versa, if we have the Hamilton Jacobi equation $\Xi(x,S_x)=const$ for an entropy functional $S(x)$, then we have the Hamilton equations for $x$ and $x^*$, see \cite{OMM-I}.

\subsubsection{Hamilton equations with dissipation}
Let us now generalize the preceding formulation of Onsager's principle to party reversible systems, where the reversible part is given by Hamilton equations. The state variables $x=(r,p)$ represent position and momentum of a particle with mass $m>0$. Energy $ E(x)=\frac{p^2}{2m} + V(r)$  consists of the kinetic energy and potential energy, and $p^*=E_p=\frac{p}{m}=v$ is the velocity (assuming temperature equal to unity). State variables $x$ are coordinates on the cotangent bundle $T^*M$, $r\in M$, and their reversible evolution is described by the Hamilton equations. Their irreversible evolution is given by a dissipation potential $\Xi(r,v)$ satisfying (\ref{Xiprop}). Altogether, the reversible and irreversible parts of the evolution constitute a GENERIC time evolution
\begin{equation}\label{Onsager2}
\left(\begin{array}{cc}\dot{r}\\ \dot{p}\end{array}\right)=\left(\begin{array}{cc}v\\ -V_r(r)-\Xi_v(r,v)\end{array}\right).
\end{equation}
How can we see Onsager's principle here?

The equation governing the time evolution of $p$ can be rewritten into the form
\begin{equation}\label{v1}
\dot{v}=m\left(-\Xi(r,v)-v V_r(r)\right)_v,
\end{equation}
which is a particular realization of (\ref{GENERIC}) with $\mathbb{G}=m$, $\mathfrak{S}=\Xi$, $X^*=v$, and $J=-V_r$. There is no entropy in this illustration, and the role of entropy is played by the energy $E$. This equation can be also seen as a dynamical generalization of Onsager's principle.

Analogical equations can be obtained for any cotangent bundle. Let us now consider the case where the $p-$variable quickly relaxes to its stationary value, $\dot{p}=0$. Equation (\ref{Onsager2}) then gives that
\begin{equation}\label{Onsager1}
\left(\Xi(r,v)+vV_r(r)\right)_v=0,
\end{equation}
which
determines the velocity $v$, and represents the Onsager variational principle with Rayleighian $\Xi(r,v)+ vV_r(r)$.
The time evolution of $r$,
\begin{equation}\label{v2}
\dot{r}=v,
\end{equation}
is then irreversible, since $v$ is determined as a solution of \eqref{Onsager1} and thus it is even with respect to the time-reversal transformation \cite{pre15,dynmaxent}. Explicitly, the evolution for $r$ becomes
\begin{equation}\label{v3}
 \dot{r}=\Xi^{\dag}_{v^{\dag}}(r,v^{\dag})|_{v^{\dag}=-V_r}
\end{equation}
where $ \Xi^{\dag}(r,v^{\dag})$ is the Legendre transformation of $\Xi(r,v)$ in the variable $v$, and $v^{\dag}=\Xi_v$. The right hand side of (\ref{v3}) is a solution of (\ref{Onsager1}).

This reduction can be seen geometrically, as a consequence of vanishing of the evolutionary part of the vector field governing $x$ on the image of a section $\gamma:x\mapsto p$ (MaxEnt) \cite{OMM-I}.

Summing up, the rate GENERIC formulation (\ref{v1}), (\ref{v2}) extends the Onsager variational principle  (\ref{Onsager1})
by putting it into the context of dynamics (\ref{Onsager2}). The extremization of the Rayleighian  $\Xi(r,v)+vV_r(r)$  can be made by following the time evolution governed by (\ref{v1}). We are  answering the question of why and when the Onsager principle is applicable. If the dissipation driven by the dissipation potential dominates the right hand side of the time evolution of $p$ in (\ref{Onsager2}), then $p$ evolves faster than $r$ and the time evolution of $(r,p)$ can be approximatively separated into the fast time evolution of $p$ followed by a slower time evolution of $r$. The fast time evolution is then the physical basis of Onsager's variational principle.
A detailed investigation of the existence of such separation, in the context of an example involving the quadratic dissipation potential, can be found in \cite{agrachev}.

\subsubsection{Full GENERIC}
Finally, the Onsager principle can be also formulated for the full GENERIC evolution, without the restriction to the Hamilton equations and cotangent bundles. Since the GENERIC evolution can be reformulated as
\begin{equation}
\dot{x} = \Psi_{x^*}
\end{equation}
with $\Psi(x,x^*) = \langle x^*, L(x) E_x\rangle + \Xi(x,x^*)$, we can carry out the Legendre transform to
\begin{equation}\label{eq.psi}
\Psi^*_{\dot{x}} = x^*,
\end{equation}
where $\Psi^*(x,\dot{x}) = \sup_{x^*} (-\Psi(x,x^*) + \langle x^*, \dot{x}\rangle)$. Equation \eqref{eq.psi} represents the Onsager's principle, and in the special case of no mechanics ($L=0$), it reduces to the pure gradient dynamics discussed in Section \ref{sec.GD}.

If the potential $\Psi$ satisfies the stationary Hamilton-Jacobi equation $\Psi(x,\Phi_x)=const$, then GENERIC is the critical state of action
\begin{equation}
\int_{t_0}^{t_1} \Psi^*(x,\dot{x})dt,
\end{equation}
and vice versa, which can be seen as a generalization of the principle of least dissipation to cases involving also mechanics \cite{OMM-I}.

Also from the Fenchel inequality \eqref{fenchel}, we get that
\begin{equation}
\Psi(x,x^*)+\Psi(x,\dot{x}) \geq \langle x^*,\dot{x}\rangle,
\end{equation}
which is related to the Onsager-Machlup variational principle \cite{om,hco-jnet2020-I}.


The full GENERIC formulation can be illustrated on an example from Section 4 in \cite{doi-onsager}. In the  notation that is adapted to this paper, Doi's example is the following. The state variable is a q-dimensional vector $x^*=(x^*_1,x^*_2,...,x^*_q)$, the Rayleighian is
\begin{equation}\label{RayDoi}
\Psi(x^*,\JJ)=\frac{1}{2}\langle x^*,\zeta x^* \rangle +\langle x^*,J\rangle
\end{equation}
where
$J=-\Gamma^Ty^*$, $\zeta$  is a positive definite matrix, $y=(y_1,y_2,...,y_p)$ is a p-dimensional  vector,  $y^*=\Phi_y(y)$, $\Phi(y)$ is the free energy, $\Gamma$ is an $p\times q$ matrix, and $\Gamma^T$ is its transpose. Equation
\begin{equation}\label{v10}
\dot{y}=\Gamma x^*
\end{equation}
then governs the time evolution of $y$.

Now we put Doi's example to the framework  of the rate GENERIC, as a reduction in the GENERIC dynamics
\begin{equation}\label{v4}
\left(\begin{array}{cc}\dot{x}\\ \dot{y}\end{array}\right)=\left(\begin{array}{cc}0&-\Gamma^T\\ \Gamma &0\end{array}\right)
\left(\begin{array}{cc}x^*\\y^*\end{array}\right)-\left(\begin{array}{cc}\zeta x^*\\0\end{array}\right).
\end{equation}
If the dissipation  in the second term on the right hand side of (\ref{v4}) is dominant, then $x$ (and also its conjugate $x^*$) evolves in time faster than $y$. The time evolution of $x^*$ is governed by (\ref{GENERIC}) with the rate thermodynamic potential that is identical to Rayleighian and with  $\mathbb{G}$ given in (\ref{rgeneric}). Onsager's principle (\ref{minrent}) determines its  asymptotically reached state $\widehat{x}^*$. The reduced time evolution is thus governed by
\begin{equation}\label{redDoi}
\dot{y}= \Gamma\widehat{x}^*.
\end{equation}
With the dissipation appearing in (\ref{v4}), we easily find that  $\widehat{x}^*=-\zeta^{-1}\Gamma^Ty^*$ and thus $\dot{y}=
-\Lambda y^*$, where $\Lambda=\Gamma \zeta^{-1} \Gamma^T$.

Summing up, by seeing Doi's  example in the context of the rate GENERIC we can answer the following questions: Why Equation (\ref{GENERIC1}) has the form (\ref{v10})? Because the first term on the right hand side of (\ref{v4}) is required to be Hamiltonian.
What is the multiscale physical basis of Onsager's principle? It is the reduction of (\ref{v4}) to (\ref{v10}) with $ x^*=\widehat{x}^*$, which is justified when the dissipation dominates the time evolution of $(x,y)$.

\subsubsection{Summary of the relation between GENERIC and Onsager's principle}
Many formulations and application of the physics related to Onsager's variational principle  have been worked out and can be found in the literature.  Different contexts in which the principle is considered and applied bring  different  physical insights. With no claim of completeness, we mention in particular the pioneer work of  Rayleigh \cite{Rayleigh}, Onsager \cite{onsager1930}, Prigogine \cite{prigogine-tip}, and Gyarmati \cite{gyarmati}. Among more recent investigations we mention  the  Rajagopal, Srinivasa, and M{\' a}lek \cite{raja-sri}, \cite{pepa-raja-karel} and their analysis of complex-fluid constitutive relations, and Guo and his collaborators \cite{thermomass} with analysis of the non-Fourier heat conduction. Moreover, new statistical insight has been gained in works by Renger, Mielke, Peletier, Montefusco, and Öttinger \cite{mielke-potential,mielke-peletier,hco-jnet2020-I} by formulating the Onsager-Machlup principle in the context of large deviations, and by Maes and Netočný \cite{maes-minep}. The rate GENERIC formulation developed in this paper contributes to all these investigation by connecting them with reductions of  dynamical theories to other dynamical theories involving less details. The connection of Onsager's variational principle with  the multiscale thermodynamics addresses both its  foundation  and its applications.

Regarding the foundation, we recall that the existence of the approach to  equilibrium states, known as the zero law of thermodynamics, provides  foundation to the classical equilibrium thermodynamics. We have  seen in Section \ref{sectiongeneric} that the equilibrium thermodynamics indeed arises by following the approach to equilibrium generated by the GENERIC equation (\ref{generic}) to its conclusion. The existence of the approach to more macroscopic dynamics (i.e. a reduced dynamics involving less details), that we can call \textit{zero law of the rate thermodynamics}, provides foundation to Onsager's variational principle. The  rate thermodynamic potential (the Rayleighian) is extremized in the course of the time evolution approaching asymptotically the reduced theory that is  autonomous and well established  (i.e. well tested with certain class of experimental observations) independently of its relation to more microscopic (i.e. involving more details) theories.

As for the applications, the established connection to reductions allows us to utilize the large experience collected in reductions of dynamical systems  \cite{pkg}, \cite{mg-multiscale}. We shall follow this path in the next section.

\subsection{Reductions of mesoscopic dynamical theories}\label{TD}

In the previous two illustrations, we have arrived at Onsager's variational principle in three steps. First, we have formulated a reduction problem (dynamics in $M$, mapping $M\rightarrow N$, and an intention to express the extracted overall features of the dynamics in $M$ as a dynamics in $N$). Next, we have reformulated it into the form of the rate GENERIC that is then shown to be a dynamic extension of the Onsager principle. In this section we investigate in more detail the first step.
How do we  cast reductions of dynamical theories  into the form of the rate GENERIC (\ref{rG})?
After  briefly recalling some of the  reduction methods, we show that an appropriate  modification of Grad's hierarchy approach to reductions has  indeed  the form of Equations (\ref{rG}).

The behavior observed in macroscopic systems, that are free from  external forces and  internal constraints,  has been found to be well described by the classical equilibrium thermodynamics. In another type of observations the same macroscopic systems appear to be composed of microscopic particles whose dynamics is governed by the classical or quantum mechanics. In addition to these  completely microscopic  and the completely macroscopic views, there are many well established mesoscopic views, as for example the view in which macroscopic systems are seen in fluid mechanics. How can the existence of different views and associated with them different theories of macroscopic systems be reconciled?
We have seen in Section \ref{sectiongeneric} that the GENERIC time evolution offers  an  answer to reductions to  the classical equilibrium thermodynamics  in which no time evolution takes place.

Reductions to theories that involve the time evolution can be divided into two groups. The first, that we shall call Chapman-Enskog-type reductions, follow the pioneer investigation \cite{chapman-cowling} of the reduction of the Boltzmann kinetic theory to fluid mechanics. The reduction is a search for a manifold $\mathcal{M}\subset M$  that is: (i) approached as $t\rightarrow \infty$,  (ii) invariant (or at least approximately invariant) in the kinetic-theory time evolution generated by the vector field $(vf)^{(M)}$,  and (iii) the vector field  $(vf)^{(M)}|_{\mathcal{M}}$ is the vector field generating the time evolution in the reduced theory. In the Chapman-Enskog analysis, the elements $x$ of $M$ are one particle distribution functions $f(\rr,\pp)\in M$ and  $(vf)^{(M)}$ is the right hand side of the Boltzmann kinetic equation. The reduced theory is fluid mechanics with $(\rho(\rr),e(\rr),\uu(\rr))\in N$, where $\rho,e,  \uu$ are fields of mass energy and momentum.

More specifically, the manifold $\mathcal{M}\subset M$ is sought as an appropriate deformation (see e.g. \cite{gk}, \cite{dynmaxent}, \cite{ehre}) of the dissipation equilibrium manifold $\mathcal{M}^{(diss)}\subset M $, that is, in the case of the Boltzmann kinetic theory, composed of local Maxwellian distribution functions (see Section \ref{sectionmaxent}).

The reductions belonging to the second group will be called hierarchy reductions. Their  point of departure is the mapping $\Pi:M\rightarrow N$, where $M$ is the state space of a theory that involves more details and that we want to reduce to a theory involving  less details.  The state space of the latter theory is $N$. The mapping $\Pi$, that  is an important  step in the search for patterns  in the phase portrait in $M$,  is thus coming in the hierarchy reduction from considerations that do not involve the dynamics in $M$. It is important to emphasize that the physical reduction is not a local transformation (induced by the mapping $\Pi$) of the vector field on $M$ to the vector field on $N$, but a result of a global analysis of trajectories on $M$. In other words, a result of the pattern recognition process in the phase portrait is generated by the vector field on $M$, reproducing its important features in the reduced vector field on $N$. It is also important to keep in mind that the physical reduction is not just a loss of information. It is a loss of details and a gain of emerging overall features.
In the Grad analysis of the hierarchy reduction of the Boltzmann kinetic equation to fluid mechanics, the elements of  $M$ are distribution functions $x\equiv f(\rr,\pp)$ and the  elements $y$ of $N$ are hydrodynamic fields $ y\equiv (\rho(\rr),e(\rr), \uu(\rr))$ of mass, energy, and momentum. The projection $\Pi$ is defined by  $f\mapsto \Pi x= (\int d\pp f, \int d\pp \frac{\pp^2}{2m} f, \int d\pp\, \pp f)$  (we put hereafter  the mass of one particle equal to one). This projection $\Pi$ comes from comparing the physical interpretations of the state variables in $M$ and $N$. Another possibility is to find the reduced vector field on $N$ by minimizing the lack-of-fit between the MaxEnt image of the reduced vector field and the original vector field on $M$ \cite{turkington,JSP2020}.

In the next step in   hierarchy reductions,   the mapping  $\Pi:M\rightarrow N$ is extended to a one-to-one mapping $\Pi^{(hierar)}:M\rightarrow M$. In the Grad analysis the extended mapping  $\Pi^{(hierar)}:M\rightarrow M $  is defined by  \\ $ f\mapsto (\int d\pp f,\int d\pp v_i f,...,\int d\pp p_{i_1}...v_{i_k} f,...)$. This mapping is indeed an extension of the mapping $\Pi$  since \\$\Pi^{(hierar)}:f\mapsto (y, \int d\pp (p_{i_1} p_{i_2}-\delta_{i_1 i_2}\frac{\pp^2}{2m})f,\int d\pp p_{i_1}p_{i_2}p_{i_3} f,...,\int d\pp p_{i_1}...p_{i_k} f,...)$.
We can see the mapping $\Pi^{(hierar)}$ as an introduction of a structure (of coordinates) into $M$. The original vector field $(vf)^{(M)}$  in $M$ (i.e. the right hand side of the Boltzmann equation in Grad's analysis) written in the coordinates provided by $\Pi^{(hierar)}$ is the Grad hierarchy. The reduction to the dynamics involving only  $y\in N$ becomes a problem of closing the hierarchy, i.e. expressing  $\int d\pp (p_{i_1} p_{i_2}-\delta_{i_1 i_2}\frac{\pp^2}{2m})f,\int d\pp p_{i_1}p_{i_2}p_{i_3} f,..., \int d\pp p_{i_1}...p_{i_k} f,...)$ in terms of $y$. Such closure is, of course, a specification of the manifold $\mathcal{M}$. The reduced dynamics taking place in the space $N$ is governed by the closed hierarchy.

Notice that for some reductions, one has that the reduced space $N$ is a subspace of the total space $M$. One interesting question at this point is to discuss algebraically the details lost in the reduction that is to investigate the quotient space $M/N$. For the case of reversible part of Boltzmann equation, one can have a projection of the first two kinetic moments of the distribution function. The reduced dynamics is then the compressible fluid motion, whereas $M/N$ has also its own dynamics determined by a Poisson bracket, called Kupershmidt-Manin bracket \cite{tronci2009,GiHoTr08}.  In a recent study \cite{esen2020matched,momentum-Euler}, it is shown that the relationship between $N$ and $M/N$ can be investigated through the matched-pair geometry, permitting also mutual interactions between $N$ and $M/N$. This strategy identifies the individual motions of $N$ and $M/N$ as subsystems of $M$ while labeling properly the rest of the terms in the dynamics in $M$ in terms of the mutual actions of $N$ and $M/N$.

The hierarchy reduction is an alternative   strategy for identifying the manifold $\mathcal{M}$. Nevertheless, the hierarchy reduction has already some elements of the  rate GENERIC. The higher moments $\int d\vv p_{i_1}...p_{i_k} f$; $k=3,,...$, that need to be expressed in terms of the first five moments,  are making their appearance in the vector fields generating the reduced time evolution. We can thus interpret the time evolution of the higher moments as the time evolution of the vector fields generating the reduced dynamics. The problem of closing the hierarchies  becomes  thus the problem of identifying  fixed points and thus technically the same problem as in reductions to equilibrium. We expect that finding fixed points that are approached as $t\rightarrow \infty$ is easier than finding invariant manifolds that are approached as $t\rightarrow \infty$.
The equations governing the time evolution of the higher moments do not have however the form of Eq.(\ref{GENERIC}). Can we modify the hierarchy reduction in such a way that the hierarchy takes the form of Equations (\ref{rG})? We shall now discuss such modification.

\subsection{Poisson   hierarchies}

We begin with the Hamiltonian time evolution in $M$ with a constant Poisson bivector $L$. Next, we  transform $L$  into a hierarchy, and only then we introduce dissipation and external forces.

For simplicity, we choose  $M$ to be a finite dimensional linear space with coordinates $x=(x_1,...,x_n)$, and the Poisson bivector $L$ with constant coordinates $L_{ij}$. The Hamiltonian time evolution equation (\ref{Hamrv}) has the form
\begin{equation}\label{Grad1}
\dot{x}_i=L_{ij}E_{x_j}
\end{equation}
where $E(x)$ is the energy.

Next, we introduce a transformation
\begin{equation}\label{Grad2}
M\rightarrow N;\,\,x\mapsto y=y(x)
\end{equation}
where $y=(y_1,...,y_m); m<n $.  We assume that the transformation (\ref{Grad2}) is linear, $y_{\alpha}=c_{\alpha j}x_j$.
Instead of applying the transformation (\ref{Grad2}) directly on the time evolution equation (\ref{Grad1}), as it is done in the Grad-like hierarchy reformulations, we apply it only on the Poisson bivector $L$. Such Poisson hierarchies  have already been introduced in \cite{mg-external} in the context of the BBGKY hierarchy and in \cite{miroslav-grad} in the context of Grad's hierarchy. In both cases the final hierarchy is an infinite hierarchy. The modification made below leads to a hierarchy composed of only two equations.

When functionals $A$ and $B$ that depend on $y(x)$ are plugged in the Poisson bracket,
\begin{equation}\label{Grad3}
\{A,B\}=A_{x_i}L_{ij}B_{x_j},
\end{equation}
the chain rule gives $A_{x_i}\rightarrow c_{\alpha i}A_{y_{\alpha}} + A_{x_i}$, and the bracket transforms to
\begin{eqnarray}\label{Grad4}
\{A,B\}&=&A_{y_{\alpha}}c_{\alpha i}L_{ij}c_{\beta j}B_{y_{\beta}}\nonumber \\
&&+A_{y_{\alpha}}c_{\alpha i}L_{ij}B_{x_j} +A_{x_i}L_{ij}c_{j\alpha}B_{y_{\alpha}}\nonumber \\
&&+A_{x_i}L_{ij}B_{x_j}.
\end{eqnarray}
The Hamiltonian time evolution equation (\ref{Grad1}) becomes another Hamiltonian equation,
\begin{subequations}\label{Grad5}
\begin{eqnarray}
\dot{y}_{\alpha}&=&c_{\alpha k}c_{\beta j}L_{kj}E_{y_{\beta}}+c_{\alpha k}L_{kj}E_{x_j}\\
\dot{x}_i&=& L_{ij}E_{x_j}+L_{ij}c_{\alpha j}E_{y_{\alpha}},
\end{eqnarray}
governing the time evolution of $(y,x)$.
\end{subequations}

From the physical point of view, the choice of  functions $A(x)$ and $B(x)$ in the Poisson bracket (\ref{Grad3}) and of the energy $E(x)$ and the entropy $S(x)$ as functions of $(y(x),x)$ amounts to   assigning to $x$ a new role.  After making the transformation, the state variable $x$ expresses  the information that cannot be expressed in terms of $y(x)$. In other words, we can see the mapping $x\rightarrow (y(x),x)$ as an introduction of a structure into the characterization of macroscopic systems. We see them as macroscopic systems  endowed with an internal structure. The overall features are characterized by $y$  and the internal structure  by $x$.

In the simplest example, the initial characterization is made by  $x=(x_1,x_2)$,  $E(x)=\frac{x_1^2}{2}+\frac{x_2^2}{2}$. The transformation (\ref{Grad2}) is defined by $y(x)=\frac{x_1+x_2}{2}$. From the physical point of view, the system under investigation is composed of two identical particles of mass equal to one, $x_1$ and $x_2$  are their velocities, $\frac{x_1^2}{2}+\frac{x_2^2}{2}$  is their kinetic energy. The transformation $x\rightarrow (y(x),x)$ changes the  viewpoint.   The system composed of two particles is after the transformation seen  as a system composed of one particle (with the velocity $y$ and kinetic energy $2\left(\frac{y^2}{2}\right)$) endowed with an internal structure characterized by the velocities $(x_1,x_2)$ that contributes to the internal energy $\frac{(x_2-x_1)^2}{4}$ and  to the  total  kinetic energy.
Still in other words, if we were using stochastic structures in our analysis, then $x$ would be a random variable, $y$ its average,  and  $x$ after the transformation would describe fluctuations around $y$.

An alternative way of arriving
at (\ref{Grad5}) is  by starting with two systems, one with the state variable
$\zeta$ and the Poisson bracket $\{A,B\}^{(1)}=A_{\zeta_{\alpha}}\mathcal{L}_{\alpha \beta}B_{\zeta_{\beta}}$ and the other with the state variable $\xi$ and the Poisson bracket $\{A,B\}^{(2)}= A_{\xi_i}L_{ij}B_{\xi_j}$. Next, we combine  both systems with the one-to-one transformation $y_{\alpha}=\zeta_{\alpha}+c_{\alpha i}\xi_i$ and $x_i=\xi_i$. The new Poisson bracket is (\ref{Grad4}) with an additional term   $A_{y_{\alpha}}\mathcal{L}_{\alpha \beta}B_{y_{\beta}}$. The Hamilton  equations governing the time evolution of $(y,x)$ are (\ref{Grad5}) with an  additional term $\mathcal{L}_{\alpha \beta}E_{y_{\beta}}$ in the equation governing the time evolution of $y$.

We now modify (\ref{Grad5}) by replacing the energy $E$ with the thermodynamic potential $\Phi$ (see (\ref{Generic})) and by introducing dissipation to the time evolution of $x$
\begin{eqnarray}\label{Grad6}
\dot{y}_{\alpha}&=&c_{\alpha k}c_{\beta j}L_{kj}\Phi_{y_{\beta}}+c_{\alpha k}L_{kj}\Phi_{x_j}\nonumber \\
\dot{x}_i&=& L_{ij}\Phi_{x_j}+L_{ij}c_{\alpha j}\Phi_{y_{\alpha}} -\Xi_{x^*}(x^*)|_{x^*=\Phi_x}.
\end{eqnarray}
In order to simplify the notation, we absorb the parameter $e^*$ in the time $t$ (i.e. the time $t$ in (\ref{Grad5}) is $e^*t$).

The second equation in (\ref{Grad6}) becomes
(\ref{GENERIC}) with
\begin{eqnarray}\label{G7}
\Psi(x^*)&=&\Xi(x^*) +x^*_iL_{ij}\Phi_{x_j} +x^*_iL_{ij}c_{\alpha j}y^*_{\alpha}\nonumber \\
\mathbb{G}_{ij}&=& Hess^{*-1}(x^*)
\end{eqnarray}
and (\ref{GENERIC1}) with
\begin{equation}\label{Grad7}
\dot{y}_{\alpha}=c_{\alpha k}c_{\beta j}L_{kj}y^*_{\beta}+c_{\alpha k}L_{kj}x^*_j.
\end{equation}
By  $Hess ^*$ we denote  the Hessian of the Legendre transformation $\Phi^*(x^*)$  of $\Phi(x)$ and $Hess^{*-1}(x^*)$ is its inverse,

In summary, we have started with a reduction problem consisting of the Hamiltonian time evolution in the state space $M$ (the Poisson bivector $L$ is assumed to be independent of $x\in M$)  and a reduction mapping $M\rightarrow N$. We have then reformulated the reduction problem into the rate GENERIC (\ref{GENERIC}) and (\ref{GENERIC1}) which then implies, as we have shown in Section \ref{OVP},   Onsager's variational principle. The relation established between the reduction of dynamics in $M$ to a reduced dynamics in $N$  and  Onsager's principle contributes to the latter by restricting the choice of quantities  $X^*, Y, \JJ$  (see (\ref{rPsi})) entering  Rayleighian (called rate thermodynamic potential in the rate thermodynamics). We have seen this relation already  in the two examples  in the previous section and we shall  see it also in the Section \ref{illus}.

\subsection{Rate thermodynamics}\label{rt}

By evaluating the thermodynamic potential $\Phi$ at the equilibrium states reached in the GENERIC time evolution as $t\rightarrow \infty$, we have arrived  in Section \ref{et} at the fundamental equilibrium thermodynamic relation (\ref{eqrelat}). Similarly, by evaluating the rate thermodynamic potential $\Psi(x^*,\JJ)$ at the rate equilibrium states $\widehat{x}^*(\JJ)$ reached in the rate GENERIC time evolution as $\rightarrow \infty$,  we obtain the fundamental rate thermodynamic relation
\begin{equation}\label{rftr}
\mathfrak{S}^{\dag}=\mathfrak{S}^{\dag}(\JJ)
\end{equation}
where $\mathfrak{S}^{\dag}(\JJ)=\Psi(\widehat{x}^*(\JJ))$.
Both relations are  the  potentials driving the approach to fixed points and evaluated at the fixed points.

Macroscopic systems at equilibrium are subjected to external influences through the thermodynamic walls described in Section \ref{et}. The parameters $T$, $\mu$ and $P$ characterizing them are not independent. The relation (\ref{eqrelat}) among $T$, $\mu$ and $P$ is inherited from the way  the systems have been prepared for the equilibrium and from the extensivity of the equilibrium state variables $(E,N,V)$ (see Section \ref{et}). In the context of the equilibrium thermodynamics, the  extensivity  plays the role of boundary conditions.

 Macroscopic systems at states at which their behaviour is well described by a mesoscopic dynamical theory are subjected to various external  and internal forces. The parameters $\JJ$ characterizing these forces play the  role that  $T$, $\mu$, and $P$ have in the equilibrium thermodynamics.
As in the equilibrium thermodynamics,  the quantities $\JJ$ are not independent, their relation (\ref{rftr}) is inherited from the way the systems have been prepared for the mesoscopic level of observations. What is missing are the boundary conditions. Only when the physics that takes place on boundaries will be taken into account, then the rate thermodynamic relation (\ref{rftr}) will become comparable in its  usefulness with  the thermodynamic relation (\ref{eqrelat}).

\section{Illustrations}\label{illus}

In hydrodynamics and  in most well established mesoscopic dynamical theories, a  good agreement with results of  experimental observations of practical interest  is achieved with GENERIC (\ref{generic}) in which the  dissipation potential is quadratic. For instance the dissipation appearing in the particular realization of (\ref{generic}) representing the classical Navier-Stokes-Fourier hydrodynamics is generated by the quadratic dissipation potential (see e.g. \cite{pkg}).
But   such simple dissipation potentials hide many interesting new features that arise in the   rate thermodynamics. A need for a more complex nonlinear dissipation has  arisen  first in chemical kinetics \cite{gw}. Our two  illustrations of the rate thermodynamics that we shall work out in this section are  therefore taken from this theory. We show that the dynamic extension of the Doi illustration of Onsager's variational principle that we have  discussed in Section \ref{OVP} can  be interpreted, after appropriately modifying the dissipation potential, as  the rate GENERIC and Onsager's variational principle leading to the mass action law  and to the Boltzmann kinetic equation.

\subsection{Rate chemical kinetics}\label{sectionchemkin}

Let $p$ components $\mathbb{A}_1,...,\mathbb{A}_p$ undergo $q$ chemical reactions
\begin{equation}\label{reac}
\mu_{1\alpha}\mathbb{A}_1+...+\mu_{p\alpha}\mathbb{A}_p\leftrightarrows \nu_{1\alpha}\mathbb{A}_1+...+\nu_{p\alpha}\mathbb{A}_p
\end{equation}
\begin{equation}
0\leftrightarrows \Gamma_{\alpha i}\mathbb{A}_i
\end{equation}
where
\begin{equation}\label{gamma}
\Gamma_{\alpha j }=\nu_{j \alpha }-\mu_{j\alpha};\,\,\,j=1,...,p;\,\,\,\alpha=1,...,q
\end{equation}
are the stoichiometric coefficients and
\begin{equation}\label{Gamma}
\Gamma=\left(\begin{array}{ccc}\Gamma_{11}&\cdots&\Gamma_{1p}\\ \vdots&\vdots&\vdots\\ \Gamma_{q1}&\cdots&\Gamma_{qp}\end{array}\right)
\end{equation}
the stoichiometric matrix. Hereafter we use the lowercase Roman   letters ($i, j, k= 1,...,p$) to label the components and the lowercase Greek letters ($\alpha,\beta=1,...,q$) to label the  reactions. We shall also use the summation convention over  repeated indices.
Throughout this paper we limit ourselves to isothermal chemical reactions.

We now turn to Doi's example discussed Section \ref{OVP}. We interpret $y$ as a vector
\begin{equation}\label{nn}
y=\nn=(n_1,...,n_p)
\end{equation}
where $n_i; \, i=1,...,p$ is the number of moles of $i$-th component. Equation (\ref{v4}) is interpreted as an  equation governing the  time evolution of $y$. This means that $\Gamma$ is the stoichiometric matrix (\ref{Gamma}) and the components of the vector $x$
\begin{equation}\label{ww}
x=\ww=(w_1,...,w_q)
\end{equation}
are fluxes ($w_{\alpha}$ is the flux of $\alpha$-th reaction, $\alpha=1,...,q$).

We now replace the rate thermodynamic potential (\ref{RayDoi})  with
\begin{equation}\label{RayDoichem}
\Psi(\ww^*,\JJ)=\Upsilon(\ww^*)+w^*_\alpha J_{\alpha}
\end{equation}
where the dissipation potential is
\begin{equation}\label{dp1}
    \Upsilon= \sum_{\alpha=1}^q \left( 2 w^*_\alpha \arcsinh\left(2w^*_\alpha/W_\alpha(\nn)\right) - W_\alpha(\nn) \sqrt{1+\left(w^*_\alpha/2W_\alpha(\nn)\right)^2}\right)
\end{equation}
with
$J_{\alpha} =-\Gamma_{\alpha i}n^*_i$.
The reduced time evolution equation (\ref{redDoi}) is the GENERIC formulation \cite{grchemkin}
\begin{equation}\label{mal3}
\dot{n}_i=-\Xi_{n^*_i}|_{n^*=\Phi_n(\nn)}
\end{equation}
of the Guldberg-Waage mass action law \cite{gw}. The dissipation potential $\Xi$ in (\ref{mal3}) is
\begin{equation}\label{GWXi}
\Xi(\nn,\XX)=W_{\alpha}(\nn)\left(e^{\frac{1}{2}X^*_{\alpha}}+e^{-\frac{1}{2}X^*_{\alpha}}-2\right),
\end{equation}
and the thermodynamic potential
\begin{equation}\label{Phicond}
\Phi(\nn)=n_{i}\ln n_{i}+Q_{i}n_{i}.
\end{equation}
The vector $\XX^*$ is the chemical affinity vector
\begin{equation}\label{XchemGW}
X^*_{\alpha}=\Gamma_{\alpha i}n^*_i.
\end{equation}
Quantities $\WW(\nn)>0$ are related to the rate coefficients of the forward and the backward reactions (see (\ref{mal2}) below)
and  $Q_i; i=1,...,p$ are constant parameters.

It can be shown that
\begin{equation}\label{GWORP}
\Upsilon^{\dag}_{(\ww^*)^{\dag}}(\nn,(\ww^*)^{\dag})|_{(\ww^*)^{\dag}=-\Gamma^T\nn^*}=\Xi_{\XX^*}(\nn,\XX^*)
\end{equation}
where $\Upsilon^{\dag}(\nn,(\ww^*)^{\dag})$ is the Legendre transformation of $\Upsilon (\nn,\ww^*)$ in $\ww^*$, $(\ww^*)^{\dag}=\Upsilon_{\ww^*}(\nn,\ww^*)$.
Let us now summarize the results.
\\

\textbf{\textit{Mass action law}}
\\

The time evolution of the number of moles $\nn=(n_1,...,n_p)$  in chemical reactions (\ref{reac}) is governed by
\begin{equation}\label{mal1}
\dot{n}_i=\Gamma_{i\alpha}w^*_{\alpha}
\end{equation}
with the Guldberg-Waage mass-action-law constitutive relation
\begin{equation}\label{mal2}
w^*_{\alpha}=\overrightarrow{k}_{\alpha} n_1^{\mu_{1\alpha}}...n_p^{\mu_{p\alpha}}-\overleftarrow{k}_{\alpha} n_1^{\nu_{1\alpha}}...n_p^{\nu_{p\alpha}},
\end{equation}
where the matrix $\Gamma$ is the stoichiometric matrix (\ref{Gamma}), and $\overrightarrow{k}_{\alpha}$ and  $\overleftarrow{k}_{\alpha}$ are the rate coefficient of the forward and backward reactions.
\\

\textbf{\textit{GENERIC mass action law}}
\\

The mass action law (\ref{mal1}), (\ref{mal2}) is a particular realization (\ref{mal3}) of the GENERIC equation (\ref{Generic}).
Direct calculations  (see \cite{grchemkin}) indeed show that (\ref{mal3}) is equivalent to the mass-action-law (\ref{mal1}),(\ref{mal2})  provided   $\WW$,  $\overrightarrow{k}_{\alpha}$,  and  $\overleftarrow{k}_{\alpha}$  are related by
\begin{eqnarray}\label{kk}
\overleftarrow{k}_{\alpha}&=&\frac{1}{2}W_{\alpha}(\nn)e^{\frac{1}{2}\gamma_{1\alpha}(Q_{1}+1)}\left(n_1^{\nu_{1\alpha}}...n_p^{\nu_{p \alpha}}n_1^{\mu_{1\alpha}}...n_p^{\mu_{p \alpha}}\right)^{\frac{1}{2}}\nonumber \\
\frac{\overleftarrow{k}_i}{\overrightarrow{k}_i}&=&e^{\gamma_{\alpha i}(Q_{\alpha} +1)}
\end{eqnarray}
and the thermodynamic potential $\Phi(\nn)$ is (\ref{Phicond}).

The formulation (\ref{mal3}) displays the multiscale nature of chemical kinetics. More detailed descriptions (see below) arise by lifting (\ref{mal3}) to iterated cotangent bundles.
\\

\textbf{\textit{Rate GENERIC mass action law}
\\}

The mass action law (\ref{mal1}), (\ref{mal2}) arises by following solutions to
\begin{equation}\label{mal4}
\left(\begin{array}{cc}\dot{\ww}\\ \dot{\nn}\end{array}\right)=\left(\begin{array}{cc}0&-\Gamma^T\\ \Gamma &0\end{array}\right)
\left(\begin{array}{cc}\ww^*\\\nn^*\end{array}\right)-\left(\begin{array}{cc}\Upsilon_{\ww^*}\\0\end{array}\right)
\end{equation}
with the dissipation potential $\Upsilon(\nn,\ww^*)$ (see (\ref{dp1})) to its conclusion.

The governing equations  of the rate GENERIC mass action law are particular realizations of (\ref{GENERIC}) and (\ref{GENERIC1}). Equation (\ref{GENERIC}) takes in chemical kinetics the form
\begin{equation}\label{mal5}
\dot{w}^*_{\alpha}=Hess^{*-1}\Psi_{\ww^*}(\ww^*,\JJ)
\end{equation}
with the rate thermodynamic potential
\begin{equation}\label{mal6}
\Psi(\ww^*,\JJ)=\Upsilon(\nn,\ww^*)+w^*_{\alpha}J_{\alpha}
\end{equation}
with $J_{\alpha}=  -\Gamma_{\alpha i}n^*_i$ and with $Hess^{*-1}$ that is the Hessian of the inverse of the Legendre transformation in the variable $\ww$ of the thermodynamic potential $\Phi(\nn,\ww)$.
Equation (\ref{GENERIC1}) takes in chemical kinetics the form (\ref{mal1}).
\\

\textbf{\textit{Onsager's variational principle for the mass action law}}
\\

If we ignore  details of the time evolution in the rate GENERIC and limit ourselves only on its final destination, then the rate GENERIC becomes Onsager's variational principle
\begin{equation}\label{mal8}
(\Upsilon(\nn,\vv)+\vv_{\alpha}J_{\alpha})_{\vv}=0.
\end{equation}
The fluxes $\widehat{\vv}(\JJ)$ obtained by solving (\ref{mal8}) determine then the mass-action-law time evolution
\begin{equation}\label{mal9}
\dot{n}_i=\Gamma_{i\alpha}v_{\alpha}.
\end{equation}
In order to simplify the notation in the formulation of Onsager's variational principle,  we have replaced  $\ww^*$ appearing in the formulation of the rate GENERIC by $\vv$. Moreover, when we also follow the established terminology used in Onsager's variational principle, the rate dissipation potential $\Psi$ can be called a Rayleighian.

One of the important advantages of the rate GENERIC extension of the mass action law (and consequently also of its formulation in terms of  Onsager's variational principle) is a possibility to include  in a simple way external forces and internal constraints. They enter as extra terms in the Rayleighian. We intend to develop  this type of applications in a future work.

\subsection{Rate kinetic theory}\label{RTT}

In this illustration we remain in chemical kinetics but we bring it to the kinetic theory of dilute gases. We restrict our analysis to homogeneous gases in which the one particle distribution function $f(\rr,\vv)$ is independent of $\rr$. Following Boltzmann's insight \cite{Boltzmann-vorlesungen},  the events in the time evolution of dilute gases that are most consequential for the overall appearance of the phase portrait  are binary collisions. This type of interactions  can be seen \cite{waldman-transport} (see also \cite{grcontmath},\cite{gk-quasi}) as  chemical reactions. Consequently, we can use chemical kinetics to investigate the time evolution of dilute gases.

In this application of chemical kinetics,
the number of
 components as well as the number of chemical reactions representing binary collisions are infinite.
The components  $\mathbb{A}$ are labeled by the particle momentum $\vv $ (with unit mass); the index $i\in\mathbb{Z}; i=1,...,p$  is  thus replaced by $\vv \in \mathbb{R}^3 $, the components   $\mathbb{A}_i; i=1,...,p$   become $\mathbb{A}(\vv), \vv\in \mathbb{R}^3$,  and the number of moles $n_i$ becomes  one particle distribution function
\begin{equation}\label{ff}
y=f(\vv).
\end{equation}
The chemical reactions  are  binary collisions. The component    $\mathbb{A}(\vv)$ enters a binary collision with  a partner  component $\mathbb{A}(\vv_1)$. The two components  $\mathbb{A}(\vv')$ and $\mathbb{A}(\vv'_1)$  are then the outcome of the collision. The fluxes
\begin{equation}\label{fluxB}
x=g(\vv,\vv',\vv_1,\vv'_1)
\end{equation}
of the collisions are four-particle distribution functions. The indistinguishability  of the particles implies two symmetries: (i) $g(\vv,\vv',\vv_1,\vv'_1)$ is symmetric  with respect to the exchange of the particle with its partner in the collision, i.e. with respect to the transformation $(\vv,\vv')\leftrightarrows (\vv_1,\vv'_1)$,  and (ii) $g(\vv,\vv',\vv_1,\vv'_1)$ is antisymmetric with respect to the exchange of particles entering and leaving the collision, i.e. with respect to the transformation $(\vv,\vv_1)\leftrightarrows (\vv',\vv'_1)$.
The momenta $(\vv,\vv',\vv_1,\vv'_1)$
are moreover constrained by the momentum and the energy conservation
\begin{eqnarray}\label{collconst}
\vv+\vv_1&=&\vv' +\vv'_1\nonumber \\
\vv^2+\vv_1^2&=&(\vv')^2+(\vv'_1)^2
\end{eqnarray}
in the collisions.

From the physical point of view, collisions are results of Newtonian mechanics. By considering them as  results of  "chemical reactions",  we are keeping only the conservations (\ref{collconst}) while ignoring all other details of mechanics, including the inertia. By adopting  the fluxes (\ref{fluxB}) in chemical reactions as independent state variables,  we are in fact bringing the inertia back at least partially.

In order to adapt the rate chemical kinetics discussed in Section \ref{sectionchemkin} to rate kinetic theory, we introduce state space $M$ with elements
$F=\left(\begin{array}{cc}f(v)\\g(v,v_1,v',v'_1)\end{array}\right)$ and the  scalar product 
\begin{equation}
\langle F,\widehat{F} \rangle =\int dv f(v)\hat{f}(v) +\int dv\int dv_1\int dv'\int dv'_1 g(v,v_1,v',v'_1)\hat{g}(v,v_1,v',v'_1). 
\end{equation}
The functions $g$ are symmetric with respect to the exchange of $(v,v')$ with $(v_1,v'_1)$ (interchanging the particles) and antisymmetric with respect to the exchange of $(v,v_1)$ with $(v',v'_1)$ (swapping pre-collision and post-collision states).
Next,
we introduce an operator $\mathbb{D}=\mathbb{D}^{(,)}\mathbb{D}^{(1)}$, where  $\mathbb{D}^{(1)}$ and $\mathbb{D}^{(,)}$  act on functions $\phi(v,v_1,v',v'_1)$ as follows:
\begin{subequations}
\begin{eqnarray}
\mathbb{D}^{(1)}\phi(v,v_1,v',v'_1)&=\frac{1}{2}\left(\phi(v,v_1,v',v'_1)+\phi(v_1,v,v'_1,v')\right)\\
\mathbb{D}^{(,)}\phi(v,v_1,v',v'_1)&=\frac{1}{2}\left(\phi(v,v_1,v',v'_1)-\phi(v',v'_1,v,v_1)\right).
\end{eqnarray}
\end{subequations}

The rate GENERIC mass action law (\ref{mal4}) becomes the rate GENERIC kinetic equation
\begin{eqnarray}\label{extB}
\left(\begin{array}{cc}\frac{\partial f(\vv)}{\partial t}\\\frac{\partial g(\vv,\vv_1,\vv',\vv'_1)}{\partial t}\end{array}\right)&=&
\mathbb{L}
\left(\begin{array}{cc}f^*(\vv)\\g^*(\ww,\ww_1,\ww',\ww'_1)\end{array}\right)
-\left(\begin{array}{cc}0\\-\frac{1}{4}\Upsilon_{g^*(\vv,\vv_1,\vv',\vv'_1)}(f,g^*)\end{array}\right)
\end{eqnarray}
with
\begin{equation}\label{extB1}
\mathbb{L}=\left(\begin{array}{cc}0&\int d\vv_1\int d\vv'\int d\vv'_1\mathbb{D}\\-\mathbb{D}&0\end{array}\right).
\end{equation}
The dissipation potential $\Upsilon(f,g^*)$ satisfies the requirements (\ref{Xiprop}) and in addition
$\Upsilon(f,g^*)=0$ if the constraint (\ref{collconst}) does not hold. Its form
\begin{equation}\label{UpsB}
    \Upsilon(f,g^*)= \int d\vv\int d\vv'\int d\vv_1\int d\vv'_1 \left( 2 g^* \arcsinh(g^*/4W(f))-8W(f) \sqrt{1+(g^*/4W(f))^2}\right)
\end{equation}
is found from the requirement that
the reduced time evolution governed by (\ref{redDoi}) is the Boltzmann kinetic equation (\ref{GdissB}) where
\begin{equation}\label{Chemkin}
\Xi(f,f^*)=\int_{\Omega^{(\ww)}} d\vv \int_{\Omega^{(\vv)}} d\ww W(f,\vv,\ww)\left[e^{X^*(f^*)/2}+e^{-X^*(f^*)/2}-2\right]
\end{equation}
with
$W>0$. $W$ equals zero if the constraints (\ref{collconst}) do not hold, and $W$ is symmetric with respect to the exchange of $\vv$ with $\vv_1$ and $\vv'$ with $\vv'_1$ and with respect to the exchange of $(\vv,\vv_1)$ with $(\vv',\vv'_1)$. The entropy $S(f)$ is the Boltzmann entropy
\begin{equation}\label{Phikin}
S(f)=-\int d\vv f(\vv)\ln f(\vv)
\end{equation}
and $f^*(\vv)=S_{f(\vv)}$.

The  operator $\mathbb{L}$  in (\ref{extB1}) plays in kinetic theory the same role as the operator 
\begin{equation}\label{Poissono}
\left(\begin{array}{cc}0&-\Gamma^T\\ \Gamma&0\end{array}\right)
\end{equation}
 in (\ref{mal4}). Direct calculations show that both these operators are skew symmetric. Moreover, since none of the bivectors depends on the state variables, their Lie derivative with respect to the Hamiltonian vector field is zero, which means that they satisfy Jacobi identity \cite{fecko}. In other words, the bivectors form Poisson brackets. 

One way to examine the Poisson bracket \eqref{Poissono} may be done in Lie algebroid framework. A Lie algebroid is a vector bundle and not necessarily asking that the dimension of the fibers and the base manifold are the same \cite{DazordSondaz,Liber96,Weins96}. The dual bundle of a Lie algebroid is a Poisson manifold. The bracket \eqref{Poissono} is an example for such a geometry. 

\textbf{\textit{Boltzmann equation in GENERIC}}
\\

The time evolution of $f(\vv)$ is governed by \cite{Boltzmann-vorlesungen}
\begin{equation}\label{kinB}
\frac{\partial f(\rr,\vv)}{\partial t}= \int d\vv'\int d\vv_1\int d\vv'_1 \mathcal{W}(f,\vv,\vv',\vv_1,\vv'_1)(f(\vv')f(\vv'_1)-f(\vv)f(\vv_1))
\end{equation}
where $\mathcal{W}(f,\vv,\vv',\vv_1,\vv'_1)=2W(f,\vv,\vv',\vv_1,\vv'_1)/\sqrt{f(\vv)f(\vv')f(\vv_1)f(\vv'_1)}$, see \cite{dissertation} for details of the calculation.
The GENERIC form \cite{grcontmath} (\ref{GdissB}) of the Boltzmann equation (\ref{kinB}) addresses the multiscale nature of its solutions.  The entropy inequality (\ref{GGprop}) points to reduced descriptions (in particular to the equilibrium thermodynamics), and  the lifts discussed in Part I \cite{OMM-I} and Section \ref{sectionRgeneric} point to more microscopic extensions.

\textbf{\textit{Rate GENERIC Boltzmann equation}}
\\

Within the rate GENERIC, we moreover obtain that
\begin{equation}\label{RGBE}
\frac{\partial g^*(\vv,\vv_1,\vv',\vv'_1)}{\partial t}=Hess^{*-1}\Psi(f,g^*)_{g^*(\vv,\vv_1,\vv',\vv'_1)}
\end{equation}
where
\begin{eqnarray}\label{PsiBol}
\Psi(f,g^*)&=&\int d\vv\int d\vv_1\int d\vv'\int d\vv'_1
(g^*(\vv,\vv_1,\vv',\vv'_1)\nonumber \\
&&\times(-f^*(\vv')-f^*(\vv'_1)+f^*(\vv) +f^*(\vv_1))
-\Upsilon(f,g^*)
\end{eqnarray}
with the dissipation potential $\Upsilon(f,g^*)$ given in (\ref{UpsB}).
\\

\textbf{\textit{Onsager's variational principle for the Boltzmann equation}}
\\

Finally, Equation (\ref{Onsager1}) reads
\begin{equation}\label{OVPBol1}
\Psi(f,g^*)_{g^*(\vv,\vv_1,\vv',\vv'_1)}=0
\end{equation}
and
Equation (\ref{Onsager2})
\begin{equation}\label{OVPBol2}
\frac{\partial f(\vv)}{\partial t}=\int d\vv_1\int d\vv'\int d\vv'_1 g^*(\vv,\vv_1,\vv',\vv'_1).
\end{equation}

\section{Concluding Remarks}

In this paper we make a passage from the maximum entropy principle (MaxEnt) to Onsager's variational principle. We begin by  providing   MaxEnt with its dynamical basis. The maximization of the entropy is made by following the GENERIC time evolution to its conclusion. From the physical point of view, the GENERIC time evolution is a mathematical formulation of the zero law of thermodynamics. The GENERIC time evolution
describes  the preparation of macroscopic systems to equilibrium states at which their behavior is found to be well described by the classical equilibrium thermodynamics.

Next, we lift the GENERIC dynamics to the  rate GENERIC dynamics that generates the  time evolution on the iterated cotangent bundle. From the physical point of view, such lift  is a passage from dynamics in the state space to dynamics in the vector fields that generate  the time evolution  in the state space. The extremization made in Onsager's variational principle is  the passage made by following the rate GENERIC time evolution to
its conclusion.  The rate GENERIC provides  thus the dynamical basis to  Onsager's variational principle. The Onsager variational principle becomes  one of the consequences of the rate GENERIC.
The main advantage of the rate GENERIC over GENERIC  is the enlargement of applicability. The rate GENERIC is applicable also in the presence of  external forces and internal constraints  that prevent approach to equilibrium and thus  make  GENERIC inapplicable.

As illustrations, we have developed  rate lifts of the mass action law and the  Boltzmann kinetic equation (and thus also their Onsager's variational formulations) as two particular realizations of the rate GENERIC.


\end{document}